%% file: SoufriereHeatBudget_review_arxiv.tex
\documentclass{article}

\usepackage{graphicx}

\usepackage{adjustbox}

\usepackage{enumerate}
\usepackage[margin=3cm, top=2.5cm, bottom=2.5cm]{geometry}
\usepackage{amsmath}
\usepackage[position=t, singlelinecheck=off, labelfont=bf]{subfig, caption}
\usepackage{natbib}
\usepackage{nicefrac}
\usepackage[mathlines, displaymath]{lineno}
\usepackage{booktabs}
\usepackage{multirow}
\usepackage{chngcntr}
\usepackage{url}
\usepackage[utf8]{inputenc}
\usepackage{graphicx}

\usepackage{siunitx}
\DeclareSIUnit\asl{a.s.l.}
\DeclareSIUnit\fps{f.p.s.}
\DeclareSIUnit\frames{frames}
\DeclareSIUnit\kgs{\kilogram\per\second}
\DeclareSIUnit\MW{\MW}
\DeclareSIUnit\Wmsq{\watt\per\metre\squared}
\DeclareSIUnit\msq{\metre\squared}

\usepackage{scrextend}
\addtokomafont{labelinglabel}{\sffamily}

\usepackage[addedmarkup=default, final]{changes} %
\definechangesauthor[color=red]{DEJ}

\begin{document}

\title{A multi-decadal view of the heat and mass budget of a volcano in unrest: La Soufrière de Guadeloupe (French West Indies)}

\author{David E. Jessop \and
  Séverine Moune \and
  Roberto Moretti \and
  Dominique Gibert \and
  Jean-Christophe Komorowski \and
  Vincent Robert \and
  Michael J. Heap \and
  Alexis Bosson \and
  Magali Bonifacie \and
  Sébastien Deroussi \and
  Céline Dessert \and
  Marina Rosas-Carbajal \and
  Arnaud Lemarchand \and
  Arnaud Burtin
}

\date{Received: date / Accepted: date}

\maketitle

Orcid IDs: \\
David E. Jessop: 0000-0003-2382-219X \\
Séverine Moune: 0000-0002-8485-0154 \\
Roberto Moretti: 0000-0003-2031-5192 \\
Jean-Christophe Komorowski: 0000-0002-6874-786X \\
Vincent Robert: 0000-0002-9016-7167 \\
Michael J. Heap: 0000-0002-4748-735X \\
Magali Bonifacie:0000-0002-4797-043X \\
Marina Rosas-Carbajal: 0000-0002-5393-0389 \\

\begin{abstract}
Particularly in the presence of a hydrothermal system, \replaced{many volcanoes output large quantities of heat}{volcanoes output great amounts of heat primarily} through the transport of water from deep within the edifice to the surface.  Thus, heat flux is a prime tool for evaluating volcanic activity and unrest.  We review the volcanic unrest at La Soufrière de Guadeloupe (French West Indies) using an airborne thermal camera survey, and in-situ measurements \added{of temperature and flow rate through temperature probes, Pitot-tube and MultiGAS measurements}.  We deduce mass and heat fluxes for the fumarolic, ground and thermal spring outputs and follow these over a period spanning \replaced{2000--2020}{20 years}.  \replaced{Our results are compared with}{We compare our results to} published data and we performed a retrospective analysis of the temporal variations in heat flux over this period using the literature data.

We find that the heat emitted by the volcano is \SI{36.5(79)}{\MW}, of which the fumarolic heat flux is dominant at \SI{28.3(68)}{\MW}.  Given a total heated area of \replaced{\num{26780}}{\num{26779}}\,\si{\msq}, this equates to a heat flux density of \SI{627(94)}{\Wmsq}, which \deleted{value} is amongst the highest established for worldwide volcanoes with hydrothermal systems, particularly for dome volcanoes.  A major change at La Soufrière de Guadeloupe, however, is the development of a widespread region of ground heating at the summit where heat output has increased from \SI{0.2(1)}{\MW} in 2010 to \SI{5.7(9)}{\MW} \replaced{in 2020}{in the present study}.  This change is concurrent with accelerating unrest at the volcano, and the emergence of \added{two} new high-flux fumaroles in recent years.  Our findings highlight the \replaced{importance of}{need for} continued and enhanced surveillance and research strategies at La Soufrière de Guadeloupe, the results of which can be used to better understand hydrothermal volcanism the world over.  

\end{abstract}

\section*{Introduction}
\label{intro}

Hydrothermal\deleted{-volcanic} systems in active island-arc \replaced{andesitic}{andesite} volcanoes are produced by the interaction of hot magmatic fluids\added{, essentially gaseous water, CO$_2$, H$_2$S and/or SO$_2$ and HCl, produced by magma degassing at depth} with \deleted{the} marine or meteoric water at shallower depths and the host-rock \citep{SIG:EncycVolc2015, HED:Nature1994}.  \added{Cooling through interaction with water (dissolution and/or absorption into deep ground waters and mixing with meteoric and sea-water) and the host-rock causes chemical species to be reduced \citep{GIG:BV1975, MOR:Elements2020}.  Hence, the geochemical profile of fluids discharged at the surface of a hydrothermal system is typically different to that at depth}.  \replaced{Hydrothermal systems}{Such volcanoes} can undergo sudden and catastrophic changes in behaviour\replaced{.  T}{ and t}wo events in recent years\replaced{, in particular, have}{have particularly} highlighted the importance of understanding all aspects of hydrothermal volcanoes and their hazardous behaviour: the September 2014 Ontake (Japan) and December 2019 Whakaari (White Island, New Zealand)  eruptions, both of which resulted in the tragic loss of human life.

\deleted{Cooling through interaction with water (absorption into deep ground waters and mixing with meteoric and sea-water) and the host-rock at hydrothermal volcanic systems strongly modifies the geochemical profile of deep fluids produced by magma degassing (essentially water, CO$_2$, H$_2$S and/or SO$_2$ and HCl, e.g. Giggenbach, 1975; Moretti and Stefansson, 2020)}
The boiling of \deleted{the formed} geothermal liquids liberates \added{a fraction of the} dissolved gases, which fractionate into the vapour phase that ascends to the surface through steam-dominated fumaroles.  \added{Partial} condensation of these vapours into ground waters may generate steam-heated waters likely to \replaced{disperse}{flow out} laterally where they can further mix with external waters and discharge as thermal springs \citep{HED:Nature1994, SIG:EncycVolc2015}.  Therefore, significant amounts of heat are emitted as the super-heated steam\added{,} generated by these interactions\added{,} rises towards the surface through networks of cracks, fissures and more porous rock within the edifice.  The super-heated steam either \replaced{condenses}{locally condensates} near the surface or escapes to the atmosphere through fumaroles \citep{CHI:JGR2001, FIS:VolcMagHydrothermGases2015, STI:IntrusionGeothermalSystems2015}.  Heat emission\deleted{s} can occur in several forms.  First, where resistance to flow is low (high permeability subsurface)\deleted{ and} the steam \replaced{may reach}{reaches} the surface without condensing and, second, where resistance to flow is high (low permeability subsurface)\deleted{ and} the steam \replaced{may condense}{condenses} near the surface.  In \replaced{the first}{this} scenario, the fumarolic output is high and significant amounts of heat and mass are transferred to the environment.  \replaced{In the second scenario}{Second}, fumarolic output is correspondingly lower and heat is brought to the surface by \replaced{forced convection and liberated to the environment by radiation and conduction}{conduction and liberated to the environment by radiation and forced convection} \citep{HAR:ThermalRemoteSensing2013, GAU:JVGR2016}.  This leads to thermal anomalies (ground heating) and small, \deleted{very }low-flux fumaroles typically distributed over quite large areas \citep[cf.][for example]{AUB:BV1984, AUB:JVGR1999, HAR:JVGR1997, HAR:JVGR2000}.  In many cases, \replaced{ground heating}{this} far exceeds the fumarolic output in terms of energy transfer \citep{MAT:JVGR2003, MAN:GRL2019}.  Due to the high heat capacity of water, direct fumarolic degassing and diffuse small fumarole/soil degassing are generally the two major components of heat loss at hydrothermal volcanic systems \citep{AUB:JVGR1999, CHI:JGR2001}.  The final component of heat transfer in hydrothermal volcanic systems is through a network of thermal springs which typically appear along the flanks or base of the system, taking advantage of structural discontinuities.  These springs discharge water, initially heated by volcanic gases, that has either condensed deep within the edifice or nearer the surface when it has come into contact with the water table \citep{FIS:VolcMagHydrothermGases2015, STI:IntrusionGeothermalSystems2015}.

Whilst the \replaced{relative}{degree and} importance of the \replaced{different heat loss mechanisms}{aforementioned components} will vary from volcano to volcano and may vary in time at a given site, volcanic heat flow in general is indicative of \citep[e.g.][]{HAR:Tectonophysics1982, LAR:BV1999, HAR:BV2009}: 
1. The state and position of the magma body;
2. The porosity/permeability of the edifice or dome;
3. The extent of infiltration of external water into the system.
As such\replaced{, spatio-temporal variations in heat flow}{ its spatio-temporal variations} are of particular importance for both monitoring and fundamental research \replaced{and allow us to greatly constrain numerical models of the magmatic and plumbing systems \citep{REN:JVGR2016}.}{, because understanding and modelling of such variations considerably narrows the domain of solutions to a set that are very similar (temporal similarity) and congruent (spatially similar) (Di Renzo et al, 2016).}

\replaced{Hydrothermal systems play}{The hydrothermal system plays} a fundamental role in providing and enhancing the physico-chemical conditions that promote rock alteration, as well as the pressurisation of hydrothermal fluids.  These  processes act as strong forcing and triggering agents on the dynamics of volcanic activity by \deleted{promoting the} mechanical\added{ly} weakening \deleted{of} edifice-forming volcanic rock \citep{POLA:Tecto2012, WYE:JVGR2014, HEAP:JVGR2015, MOR:BV2019} and, therefore, \added{promoting} recurrent partial flank collapses \citep{LOP:Science1993, deV:Geology2000, REID:Geology2001, REID:Geology2004, JOHN:JVGR2008}, as observed at La Soufrière de Guadeloupe \citep{KOM:Gwada2005, ROS:SciRep2016}.  Escalating pressurisation of hydrothermal systems\added{,} as a result of permeability loss due to hydrothermal alteration\added{,} can also lead to explosive activity \citep{HEAP:NatCommun2019} that can reach paroxysmal levels with non-magmatic laterally-directed turbulent pyroclastic density currents or blasts (e.g. Bandaisan, Japan, in 1889).  Hydrothermal alteration has also been observed to reduce the thermal conductivity and thermal diffusivity of andesite for a given porosity \citep{HEAP:JVGR2020}.  Finally, the hydrothermal system is a strong modulator of geophysical and geochemical signals of magmatic unrest and can generate a plethora of \replaced{non-magmatic unrest}{unrest non-magmatic} signals that render monitoring, as well \added{as} the\added{ir} interpretation \deleted{of the complexity of coupled processes} and forecasting\deleted{ of their evolution towards eruptive unrest}, very challenging \citep{POU:BV2015}.

In this paper \added{which spans the past 20 years with particular emphasis on the 2010--2020 period}, we concentrate on the use of thermal measurements to infer the state of unrest of a major hydrothermal volcanic system, that of La Soufrière de Guadeloupe (Lesser Antilles).  \added{We present the first study for this volcano that fully integrates measurements of all the heat sources over such a long period of time.  }La Soufrière de Guadeloupe is a\replaced{ good}{n ideal} target for such a study due to the wealth of geochemical, geological and geophysical data acquired on the volcano.  As such\added{,} it is often considered a natural laboratory \replaced{representative of}{for} andesitic hydrothermal \replaced{systems}{volcanoes}.

%
\section*{Context}

La Soufrière de Guadeloupe (\SI{16.0446}{\degree} N, \replaced{-}{--}\SI{61.6642}{\degree} E, alt. \SI{1467}{\metre}\added{, hereby referred to as La Soufrière}) is an andesitic \deleted{type} dome volcano situated in the south of the Basse-Terre island of Guadeloupe (French West Indies), which is part of the Lesser Antilles volcanic arc and is the most recent edifice of the Grande Découverte complex (445 ka).  La Soufrière\deleted{ de Guadeloupe} is amongst the most active and potentially deadly of the volcanoes in the Lesser Antilles Arc \citep{KOM:Gwada2005}.  Hydrothermal activity is sustained by gas and heat transfer from a 6--7 km deep andesitic magma reservoir to shallower aquifers \citep{PIC:JP2018}.  Owing to an extensive hydrothermal system, La Soufrière\deleted{ de Guadeloupe} has undergone a series of six phreatic and hydrothermal explosive eruptions \deleted{(Komorowski et al., 2005)} since the last major magmatic eruption in 1530 C.E. \added{\citep{KOM:Gwada2005}}.  The \replaced{most recent}{last}, and probably most famous, eruption was in 1976--77 \citep{FEU:JVGR1983, HIN:JAV2014}.

\begin{figure*}[ht]
  \centering
  \rotatebox{-90}{
    \includegraphics[width=.85\textheight]{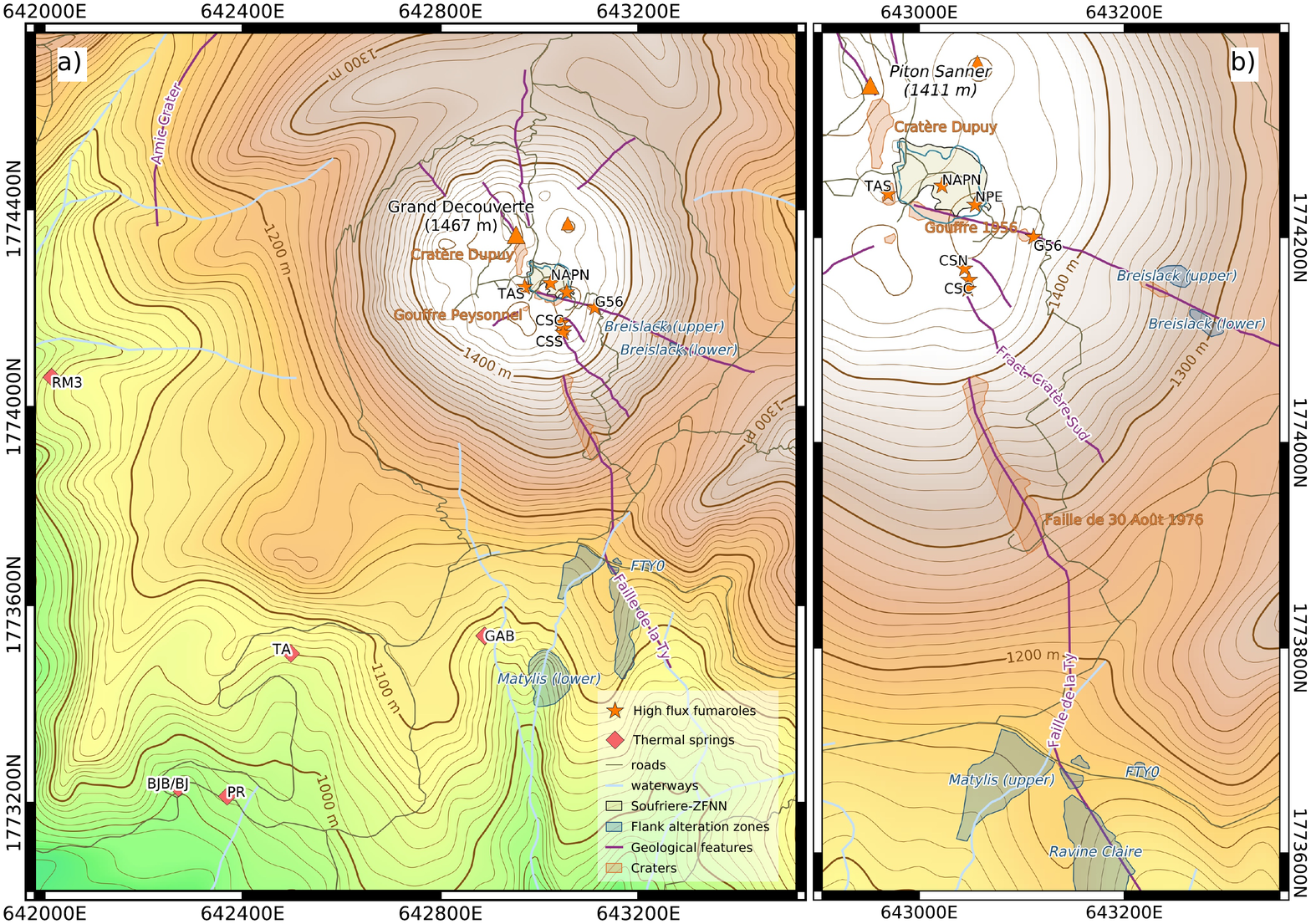}
  }
  \caption{a) Topography of La Soufrière region showing the sites (thermal
    springs and fumaroles) reported here.
    b) zoom of the summit area.  Site codes are: Cratère Sud Centre (CSC),
    Cratère Sud Nord (CSN), Cratère Sud Sud (CSS), Gouffre 56 (G56),
    Tarissan (TAS), Napoléon Est (NPE), Napoléon Nord (NAPN);
    Carbet-Echelle (CE), Galion (GA), Tarade (TA), Pas du Roy
    (PR), Bains Jaunes (BJ) and Ravine Marchand (RM3).  The base DEM,
    hill shading and contours were calculated from an aerial LiDAR survey at
    5 m resolution.  Road, footpath and waterway information were obtained
    via the French Government’s open data platform (
    https://www.data.gouv.fr/, accessed 2020-04-16).  Geological
    information is as presented in \cite{LES:GJI2012, BRO:JVGR2014,
      MOR:JVGR2020} and references therein.}
  \label{fig:soufriere_overview}       
\end{figure*}

The present edifice dates back \deleted{to} at least 9150 years \citep{KOM:Gwada2005, LEG:ScenariiEruptifsLaSoufriere2012}, during which time several major magmatic eruptions have occurred, the latter in around 1530 C.E.\replaced{, when}{ at which time} the current dome was emplaced \citep{KOM:Gwada2005, BOU:JVGR2008}.  Since this last magmatic event, there ha\replaced{ve}{s} been a number of phreatic and/or hydrothermal explosive eruptions.  The last eruption occurred in 1976--77, following which the volcano became essentially dormant until 1992 when seismic activity and steam emissions from summit fumaroles recommenced \citep[OVSG-IPGP 1999-2001\footnote{http://www.ipgp.fr/fr/ovsg/bulletins-mensuels-de-lovsg};][]{ZLO:JVGR1992, KOM:ActiviteEruptiveSoufriere2001, KOM:Gwada2005}.  Summit degassing has gradually increased concomitantly with other observables (seismic, gas flux and concentration, ground and fumarole temperatures, deformation, emissions of chlorine-rich acid gases)\added{,} over the past $\sim30$\,years\replaced{.  This has included the appearance}{ including the apparition} of two new high-flux fumaroles \citep[Napoléon Nord and Napoléon Est, labelled NAPN and NPE on Fig.\:\ref{fig:soufriere_overview}; OVSG-IPGP 2014-2016;][]{KOM:Gwada2005, VIL:JVGR2014, MOR:JVGR2020}, extensive zones of substantial surface heating and \deleted{``}scalding\deleted{''} of \deleted{the} vegetation.  Several fumarolic sites on the flanks characterised by a low state of activity since 1976\added{,} gradually vanished\replaced{.  At the summit,}{ with} Tarissan (TAS), Cratère Sud (CS), la Fente du Nord, Gouffre 56 (G56) and the Lacroix \replaced{fumaroles}{fumerolles} \replaced{had all become}{all becoming} inactive by 1984 \citep{KOM:Gwada2005, BOI:JVGR2011, FEU:Soufriere2011, RUZ:CG2013}.

A\replaced{n increase}{ spike} in activity in 2018 \replaced{raised}{increased} speculation that the volcano is in a state of growing unrest and is likely to undergo another eruptive episode in the near future \citep{MOR:JVGR2020}.  \deleted{This uncertainty as to the ongoing evolution of the volcano is further evidenced by the fact that,} Until 2014, ground thermal anomalies and accompanying soil degassing had likely been limited to the areas directly surrounding the major fumaroles, as well as the Faille de la Ty/Ravine Claire/Matylis structure \citep[Fig.\:\ref{fig:soufriere_overview}; OVSG-IPGP 2014-2020;][]{KOM:Gwada2005, LES:GJI2012, BRO:JVGR2014}.  In recent years, however, a number of thermal anomalies and altered zones have been observed such as at the Zone Fumerolienne Napoléon Nord (ZFNN) at the summit, delimited by NAPN, Cratère Dupuy (DUP) and TAS, adjoining the Breislack fault (BLK) and in the upper Matylis ravine (Fig.\:\ref{fig:soufriere_overview}, OVSG-IPGP 2014-2020 and this work).  \replaced{I}{Relatedly, i}ncreasing fluxes and acidification of the water and gas rising up within the volcano has led to significant alteration and weakening of the edifice, leaving it vulnerable to flank collapse during even moderate seismic activity or extreme rainfall \citep{KOM:Gwada2005, ROS:SciRep2016}.

The summit vents are located near major fractures and fault zones, i.e. zones of high vertical permeability \citep[\added{see Fig.\:\ref{fig:soufriere_overview} and}][]{ZLO:JVGR1992, KOM:Gwada2005}.  These are likely to have acted as a \replaced{magma ascent route for}{route of ascent for the magma that formed} the dome \citep{BRO:BV2000}\replaced{.  T}{, t}he deepest part of \replaced{these fractures}{which} act\deleted{s} as a zone of preferential input of magmatic gases into the hydrothermal aquifer, and its shallowest part behaves as a zone of preferential discharge for the hydrothermal aquifer \citep{BRO:BV2000}.  The horse-shoe shaped scar of recurrent partial edifice collapses \replaced{over}{of} at least the last 3000 years, including the major Amic Crater (1370 BCE) and the 1530 CE events, form\added{s} a listric clay-rich low-permeability south-sloping surface\replaced{.  This allows}{ for} preferential outflow of \replaced{heated ground waters through}{groundwaters that were heated.  This led to the emergence of} a number of thermal springs \citep{BRO:BV2000, VIL:EPSL2005, RUZ:CG2013, VIL:JVGR2014}.  Here, fluids are heated within the hydrothermal system and then \added{partially} cooled \deleted{to some degree} by mixing with meteoric water before escaping to the environment.

%
\section*{Materials and methods}

\replaced{Aerial thermal surveys were carried out in 2010 and 2019, MultiGAS and Pitot-tube measurements have been carried out monthly since 2017 and the thermal springs have been sampled monthly since 2000}{Here  we  present  the  first  study  that  fully  integrates  measurements  of  all  the  heat  sources  that span over the same time window (typically monthly, from 2000-present day).  Therefore, they are contemporaneous even if sampling rates are different between different methods and sites}.  Our measurements are effectively contemporaneous even though sampling times differ between different methods and sites.  \added{We calculated errors on our estimations using standard error propagation formulae \citep{KU:NBS1966, GIB:TheSystematicExperiment1986}.  Examples of how to apply these formulae and a table of relative standard errors for all the parameters used in this study can be found in the Supplementary Material.}

\subsection*{Ground thermal anomaly flux}

We used airborne thermal imagery to measure the extent and distribution of \deleted{these} thermal anomalies over the entire volcano\replaced{ using an InfraTec VarioCam HD thermal camera (8--\SI{14}{\um}) with $640 \times 480$ pixel resolution.  A 15 mm focal length lens ($56.1 \times 43.6$\si{\degree} FOV), gave an instantaneous field of view (IFOV) of \SI{1.65}{\milli\radian}}{.  The thermal camera used was an InfraTec VarioCam HD with $640 \times 480$ pixels resolution which, combined with a 15 mm focal length lens ($56.1 \times 43.6$\si{\degree} FOV), gives an instantaneous field of view (IFOV) of 1.65 mrad}.  \replaced{The distinguishable temperature difference between neighbouring pixels, $\mathrm{NE}\Delta T$, was \SI{0.5}{\kelvin}}{The noise equivalent temperature difference, $\mathrm{NE}\Delta T$, was 0.5 K, meaning that the temperature difference between neighbouring pixels needs to be greater than 0.5 K to be distinguishable}.  We used an Isotech Calisto calibration oven with black-body source to calculate the drift of the camera's temperature measurements and applied this to our thermal images.  

The airborne thermal survey was conducted on 22 November 2019 with helicopter support provided by the local Civil Protection Service and in pre-dawn conditions \replaced{(05:40--06:05 local time, sunrise was at 06:14 local time)}{The first images were taken at about 05:40 local time and the survey was completed before sunrise (06:14 local time)}.  \replaced{Ground heating due to incoming solar radiation was hence minimised}{The advantage of pre-dawn conditions is that ground heating due to incoming solar radiation is minimised}.  GPS locations were recorded at \SI{1}{\hertz} using a Garmin 64st \deleted{for the duration of the flight}.  Images were acquired through the open door of the helicopter from heights of about 50\replaced{--}{-}300 m \deleted{above the ground}.  \replaced{The sky was cloudless}{Weather during the flight was good with exceptional visibility (0\% cloud cover)} with only a very light wind from the North (cf. the predominant trade winds, les Alizées, blow from the East).  Little rain had fallen in the week prior to the survey\added{,} so the ground surface was dry.

We georeferenced and orthorectified our thermal images using \replaced{a}{points located on a hill-shaded} DEM calculated from Institut Geographique National (IGN) aerial photography, processed using \replaced{MicMac}{the MicMac photogrammetry software} \citep{RUP:OGDSS2017} and \added{IGN} orthophotos \deleted{provided directly by the IGN} (Fig.\:\ref{fig:georeferenced_thermal}).  Georeferencing \added{of the thermal images} was performed in QGIS using a thin-plate spline transform when the images were taken obliquely, or a Helmert transform for vertically-oriented images.  Pixel to physical distance conversions were computed as per \cite{BOM:JVGR2018}.

A schematic of the various fluxes seen by the camera is shown in Fig.\:\ref{fig:stock_flow}.
The effective brightness temperature \deleted{, $T_b$,} is a function of the incoming fluxes which are functions of the temperature of the objects in the field of view through the Stefan-Boltzmann law \deleted{, $P$}.  The brightness temperature is also affected by reflection of incoming radiation (e.g. $L_{\mathrm{sol}}$ and $L_{\mathrm{atm}}$ in Fig.\:\ref{fig:stock_flow}).  T\added{hus t}he \replaced{true}{absolute, or kinetic} temperature of the ground can \deleted{thus} be expressed as

\begin{linenomath*}
  \begin{equation}
  T = \left( \dfrac{T_{\mathrm{cam}}^4 - T_{\mathrm{atm}}^4 - (1 - \tau) T_g^4}{\epsilon \tau} \right)^{1/4},
  \label{eq:kinetic_temperature}
  \end{equation}
\end{linenomath*}

\noindent where \added{$T_{\mathrm{cam}}$ is the brightness temperature seen by the camera, }$T_{\mathrm{atm}}$ is the brightness temperature of the upper atmosphere, \added{$T_g$ is the temperature of gases between the object and camera}, \replaced{$\tau$}{$\tau_g$} is the transmissivity of an atmospheric and volcanogenic gas mixture between camera and the ground and $\epsilon$ is the emissivity of the ground (Fig.\:\ref{fig:stock_flow}).

We converted at-camera (brightness) temperature to absolute temperature by applying Eq.\:\ref{eq:kinetic_temperature}.  Fumarole plumes and areas outside the region of interest were masked.  We calculated $\tau$ \added{using a radiative transfer model}, with the surface-camera distance given by the georeferenced images and GPS location of the camera \citep{KOC:JQSRT2016, BERK:JQSRT2017}.  We took the surface emissivity to be constant for all the heated areas with $\epsilon = 0.95$ in line with that found for other studies on andesitic systems \citep{SEK:JGR1974, GAU:JVGR2016}.

We note that\replaced{ not all}{, although not 100\%} of the steam condenses before reaching the surface.  \replaced{Condensed liquids drain away to be discharged elsewhere in the system (i.e. through thermal springs, in which case the heat transported is accounted for in the thermal springs heat budget) and any residual heat transferred to the ground, where it is accounted for in the soil heat budget.}{Liquid water formed through condensation typically does not reach the surface and either drains away to be output elsewhere in the system (i.e. through thermal springs) or transfers its heat to the ground.}  Hence we do not consider heat transported by condensed water here \citep[cf.][]{GAU:JVGR2015}.  Our heat balance is thus \citep{SEK:JGR1974, MAT:JVGR2003, HAR:ThermalRemoteSensing2013, MAN:GRL2019}
\begin{linenomath*}\begin{align}
  Q_{\mathrm{soil}} & = Q_{\mathrm{soil, rad}} + Q_{\mathrm{so\added{i}l, conv}} \\
  Q_{\mathrm{soil, rad}} & = \added{A_{\mathrm{heated}}} \epsilon_{\mathrm{soil}} \sigma \left( T^4 - T^4_{\mathrm{amb}} \right) \\
  Q_{\mathrm{soil, conv}} & = A\added{_{\mathrm{heated}}} h_c \left( T - T_{\mathrm{amb}} \right)
  \label{eq:ground_fluxes}
\end{align}\end{linenomath*}

\noindent where \added{$Q_{\mathrm{soil}}$ is the soil heat flux and subscripts $\mathrm{rad}$ and $\mathrm{conv}$ refer to radiative and conductive components of $Q_{\mathrm{soil}}$, respectively,} $T$ is the ground temperature, \added{$T_{\mathrm{atm}}$ is the ambient temperature and $A_{\mathrm{heated}}$} is the heated area, $\epsilon_{\mathrm{soil}}$ is the soil emissivity.  The heat transfer coefficient, $h_c$, depends on several factors, particularly the local wind speed, $w$.  We use the Schlichting-Neri model (\citealp{NERI:JVGR1998}; \added{\citealp{GAU:JVGR2013}})
\begin{linenomath*}
  \begin{equation}
  h_c = 1500 w(z) \left( 1.89 + 1.62 \log(z/z_0) \right)^{-2.5},
  \label{eq:schlichting_neri}
  \end{equation}
\end{linenomath*}

\noindent where $z$ is the height above the surface and $z_0$ is a measure of the surface roughness.  Eq. \ref{eq:schlichting_neri} has been shown to produce results that are consistent with the surface heat balance at La Soufrière\deleted{ de Guadeloupe} \citep{GAU:JVGR2013}, such that the heat conducted to the surface equals $Q_\mathrm{soil}$.  We note that the surfaces on the volcano where heat transfer occurs consist typically of centimetric blocks and thus we take $z_0 = $\:\SI{0.01}{\metre} as our roughness scale.  We determined $w$ from measurements at the Sanner weather station (\added{cf.~Piton Sanner in} Fig.\:\ref{fig:soufriere_overview}) at the time of thermal image acquisition.  The anemometer at Sanner is approximately \SI{2}{\metre} above ground level, so we take $z = $\:\SI{2}{\metre} in our calculations.  For wind speeds between 5--\SI{10}{\metre\per\second}, as seen on the 22 November, we find $h_c$ between 21.1 and \SI{42.3}{\Wmsq\per\kelvin}.  Considering error propagation, we estimate a relative standard error of about 10\% on the radiative and convective flux measurements, and thus about 15\% for the total flux.

\subsection*{Fumarole heat and mass fluxes}

\subsubsection*{In-plume fumarole steam flux via MultiGAS traverses}
The OVSG\footnote{Observatoire Volcanologique et Simologique de la Guadeloupe} MultiGAS consists of an IR spectrometer for CO$_2$ determination and electro-chemical sensors for SO$_2$, H$_2$S and H$_2$.  The atmospheric pressure \deleted{(Patm)} is determined with the sensor installed on the CO$_2$ spectrometer card.  The MultiGAS also includes an externally-fitted relative humidity (RH) sensor (Galltec, range: 0--100\% RH, accuracy: $\pm2$\%) and temperature sensor (range: \deleted{[}-30\added{--}70\deleted{]}\si{\celsius}, resolution: \SI{0.01}{\celsius}), \replaced{to determine water vapour concentration \citep{MOU:JVGR2017}}{that can be used to determine the concentration of water vapour following the procedure described by Moussallam et al. (2017)}.  H$_2$O determination with these external sensors \replaced{reduced the risk of underestimating the measured water/gas ratios due to steam condensation in the inlet}{allowed us to circumvent the potential influence of steam condensation in the MultiGAS inlet tubing and, therefore, to avoid underestimating the measured water/gas ratios}.  An onboard GPS receiver tracked the location of the instrument at \SI{1}{\hertz}\replaced{.  D}{ and the d}ata were visualised on an external tablet \replaced{in real time}{connected in real-time via wifi}.  More detailed information about the OVSG MultiGAS, its design and performance characteristics can be found in \cite{TAM:Geosci2019} and \cite{MOR:JVGR2020, MOR:IJG2020}.  

Fumarolic gas fluxes \replaced{we}{a}re determined for the three main vents that generate plumes (CS, TAS and G56, Figure 1) following \deleted{the methodology initially laid out by} \cite{ALL:CG2014} and \deleted{improved upon by} \cite{TAM:Geosci2019}.  The horizontal and vertical distributions of gas species in the plume cross-sections were measured \added{a few meters downwind from the vents} during \replaced{orthogonal traverses on foot}{walking traverses orthogonal to the plume direction}\deleted{, a few meters downwind from the vents}.  Gas concentrations \replaced{we}{a}re measured at two different heights (typically 0.9 and \SI{2}{\metre}) as \deleted{, during most past and current measurements, }the volcanic gas plumes are \added{generally} flattened to the ground by strong trade winds (2--\SI{14}{\metre\per\second}) and have a maximum height of ca. 3--\SI{4}{\metre} above the ground at each measuring site with a maximum gas density centred at between $\sim1.5$ and \SI{2}{\metre} above the ground \citep[our visual observations;][]{GAU:JVGR2016, TAM:Geosci2019}.  For each site, we interpolated the concentration measurements using a 2D spline function and then integrated over the plume cross section to obtain integrated concentration amounts (ICAs) using RatioCalc \deleted{software} \citep{TAM:CG2015}.  The CO$_2$ fluxes are derived by multiplying the CO$_2$ concentration integrated over the plume cross section with the wind speed measured during the gas survey with a hand-held anemometer.  We use CO$_2$ as the volcanic marker\deleted{ to avoid any flux underestimation} as\deleted{ it has been shown that}, due to its more conservative behaviour compared to H$_2$S and due to the faster response of the IR CO$_2$ sensor compared to the \added{electro-chemical} H$_2$S\deleted{ chemical} sensor, \replaced{this way the MultiGAS is }{CO$_2$ sensors are} able to detect rapid concentration changes during plume transects\replaced{.  This avoids flux underestimations and}{ which} leads to more accurate gas flux measurements \citep{TAM:Geosci2019}.  Due to the high atmospheric background for H$_2$O and CO$_2$, our walking profiles start and end in pure atmospheric background in order to characterise and then subtract the ambient air composition from our recorded data.  Steam fluxes, $\dot{m}$, are derived from the CO$_2$ flux by multiplying it by the weight ratio of H$_2$O/CO$_2$.  Steam flux estimates were possible only when water was successfully determined via the external RH sensor.  \replaced{It is important to note that some temporal variability of steam fluxes could be due to: i) different ambient humidity and weather condition at the summit between field measurements; ii) occasional partial steam condensation on the external sensors}{It is important to note that some variability of steam fluxes could be due to both high ambient humidity on top of the volcano (RH close to 100\%), occasional partial steam condensation on the external sensors and rapid weather condition changes at the summit}.  Indeed, particularly for tropical volcanoes such as La Soufrière\deleted{ de Guadeloupe}, water vapour in the plume rapidly condenses upon contact with the atmosphere.  However, this condensed water is not taken into account by the MultiGAS measurements.  It has been shown that, in such tropical conditions, properly accounting for the condensed water adds approximately 35\% to the steam flux estimations \citep{GAU:JVGR2016}, an increase which we consider in our analysis.  Lastly, wind speed is the main source of error in quantifying volcanic gas fluxes, leading to typical standard errors on steam flux estimation of about 40\%.

\subsubsection*{At-vent fumarole fluxes via Pitot-tube measurements}
Measurements of the steam exit speed at the vent of several fumaroles were made using a Pitot-tube instrument based around Freescale MPX2200AP and MPX2010DP temperature compensated pressure sensors that measured the dynamic pressure in the moving stream and ambient (stagnation) pressure.  Pressure readings were taken at \SI{3.75}{\hertz} and the median of 10 measurements was recorded by an Arduino Due.  Uncertainty in the pressure readings was \SI{3.1}{\pascal}, meaning that the minimum \replaced{detectable}{recordable} speed was about \SI{5}{\metre\per\second}.  From these values, the speed of a moving stream of gas, $u$, of density $\rho$ was calculated as \citep{MAS:MechFluids1998}
\begin{equation}
  u = \sqrt{ \dfrac{2 \Delta p}{\rho} }
  \label{eq:pitot_speed}
\end{equation}

\noindent where $\Delta p = p_0 - p$ is the dynamic pressure, $p_0$ is the stagnation pressure and $p$ is the free-stream pressure.  Vent temperatures were simultaneously measured using a PT1000 resistance temperature sensor with an instrumental error of $\pm$\SI{1}{\kelvin}.  These measurements, along with the pressure readings, were used to calculate the steam density, $\rho(p, T)$, using numerical codes based on \replaced{IAPWS thermodynamic calculations}{the IAPWS formulation which calculates the Helmholtz energy as a function of temperature and density} \citep{WAG:JPCRD1993, WAG:JPCRD2002}.  Measurements were taken repeatedly at different points across the vent in order to build up an idea of the velocity distribution.  \replaced{Typically 7 measurements were taken and the median velocity from these measurements was used in the calculations that follow}{The calculations that follow are based on the median velocity from these measurements}.

From vent speed, we deduce the mass flux from the fumaroles which, as water vapour contributes up to 98\% of the total mass \citep[][OVSG-IPGP bulletins 2017-2020;]{ALL:CG2014, TAM:Geosci2019, MOR:JVGR2020}, is equivalent to the steam flux,
\begin{equation}
  \dot{m} = \rho \bar{u} A \approx \rho_\mathrm{steam}(T) \bar{u} A
  \label{eq:mdot_pitot}
\end{equation}

\noindent where \added{$\bar{u}$ is the mean vent speed (equivalent in this case to median vent speed)}, $A$ is the area of the vent.  Whereas in \cite{MOR:JVGR2020}\added{,} vent area was estimated by eye by the Pitot-tube operator, here we calculate $A$ by analysing thermal images.  We repeatedly took thermal images looking straight into the vents throughout the period whe\replaced{n}{re} Pitot-tube measurements were made, from which we manually traced around the vent perimeter and, using the on-camera laser distance measurements, then converted to physical area (i.e.~in \replaced{\si{\msq}}{m2}) via a pixel-to-physical length conversion as per \cite{BOM:JVGR2018}.  We estimated the relative standard error on mass flux measurements to be 10\%.

\subsection*{Heat flux estimations}

The heat released through fumarolic activity is essentially due to cooling and condensation of the volcanic steam.  Fumarole heat flux can generally be decomposed into two contributing factors: radiation by the heated vent surface, $Q_\mathrm{rad}$, and the specific and latent heat carried by the gas phase, $Q_\mathrm{gas}$, so that $Q_\mathrm{fumarole} = Q_\mathrm{rad} + Q_\mathrm{gas}$ \citep{HAR:ThermalRemoteSensing2013, GAU:JVGR2016}.  Heat lost to the surroundings through the walls of the fumarole pipes is not considered as part of this heat budget, but are accounted for through the geothermal heating of the surrounding ground \citep{STE:JVGR1993, MAN:GRL2019}, as shown in the previous section.  Following \cite{HAR:ThermalRemoteSensing2013, ALL:CG2014, GAU:JVGR2016} we write these as
\begin{linenomath*}
  \begin{align}
  Q_{\mathrm{rad}} & = A \epsilon \sigma \left( T^4 - T^4_{\mathrm{amb}} \right)
    \label{eq:Q_rad} \\
  Q_{\mathrm{gas}} & = \dot{m} \left(c_{p,v}(T)\left(T - T_{\mathrm{boil}}\right) + L(T) + c_{p,l}\left(T_{\mathrm{boil}} - T_{\mathrm{amb}}\right) \right)
  \label{eq:Q_gas}
  \end{align}
\end{linenomath*}

\noindent where $\epsilon$ is the ground emissivity, $\sigma$ is the Stefan-Boltzmann constant, $c_p$ is the specific heat capacity \deleted{of the gas phase}, $L \approx$\:\SI{2260}{\kilo\joule\per\kg\per\kelvin} the latent heat of condensation, $T$ is the temperature of the steam, \replaced{$T_{\mathrm{boil}}$}{$T_b$}\:$\approx$\:\SI{96.7}{\celsius} is the boiling temperature of water at the dome altitude and $T_{\mathrm{amb}} \approx$\:\SI{17}{\celsius} is the ambient temperature at the summit.  The subscripts $v$ and $l$ refer to the vapour and liquid states of water, respectively, with $c_{p,v} \approx$\:\SI{2.015}{\kilo\joule\per\kg\per\kelvin} and $c_{p,l} \approx$\:\SI{4.200}{\kilo\joule\per\kg\per\kelvin} for summit temperatures and pressures.  During the survey period, $T$ ranged from 96.9 and \SI{108.6}{\celsius} for CS and has been measured in the water lake at TAS to be approximately \SI{97.5}{\celsius} (OVSG-IPGP 2016-2020).  Since \added{the G56 vent is in a c.\,\SI{30}{\m} deep cavity within the volcano, requiring specialised equipment to access,} it is \replaced{impractical}{impossible} to measure it directly\replaced{.  Thus}{,} we estimate that the temperature at the G56 vent is at the boiling temperature of water.  We note that $c_p$ and $L$ are functions of $p$ and $T$ and were solved for using similar numerical routines as for density.  $T_{\mathrm{boil}}$ is a function of pressure only and is also deduced from the IAPWS formulations \citep{WAG:JPCRD1993, WAG:JPCRD2002}.

Given the instrumental and measurement errors summarised in the text above, and using error propagation techniques \citep[see][for example]{GIB:TheSystematicExperiment1986}, we estimated the standard errors on the flux estimation using the Pitot-tube and MultiGAS instruments.  In the case of the Pitot-tube instrument, the standard error in estimating $Q_{\mathrm{fumarole}}$ is dominated by the mass flux and radiative flux terms and, overall, is \replaced{around}{of the order of} 10\%.  The standard error for our estimates based on MultiGAS measurements is dominated by the uncertainty in the mass flux measurements alone and so is about 40\%.

\subsection*{Thermal Springs}

The nine thermal springs situated around the base of the current dome have been monitored regularly by the OVSG since 1978 by manually measuring temperature and flow rate.  The majority of sites have been visited on a 1\replaced{--}{-}3 month basis to take manual temperature readings as well as physico-chemical parameters such as pH and conductivity, and to take samples for future chemical analysis \citep{VIL:EPSL2005, VIL:JVGR2014}.  During these outings, and when it was possible, volumetric flow rate, $\dot{V}$, was deduced from the time taken to fill a container of known volume.  This process was repeated 6\replaced{--}{-}10 times and we report here the mean value of these measurements.  From this, we calculate the mass flow rate, $\dot{m}_{\mathrm{spring}} = \rho \dot{V}$, which then allows us to calculate the heat flux as the sum of specific, evaporative and radiative heats,
\begin{linenomath*}
  \begin{equation}
    Q_\mathrm{spring} = Q_{\mathrm{spec}} + Q_{\mathrm{evap}} + Q_{\mathrm{rad}}
    \approx Q_{\mathrm{spec}} = \dot{m} c_{p,l}(T) (T - T_{\mathrm{amb}}).
    \label{eq:spring_fluxes}  
  \end{equation}
\end{linenomath*}

\noindent \replaced{We drop evapotransport, $Q_{\mathrm{evap}}$, and radiative heat losses, $Q_{\mathrm{rad}}$, in Eq.~(\ref{eq:spring_fluxes}) as these contribute negligibly to the heat budget}{and $ Q_\mathrm{rad} $ as per Eq.~(\ref{eq:Q_rad}).  We note that evapotransport and radiative heat losses contribute negligibly to the heat budget of the thermal springs and so are not included here}.  \added{The relative standard error on these measurements is about 5\%.}

%
\section*{Results}

\subsection*{Ground heat flux}

\begin{figure*}[ht]
  \rotatebox{-90}{
    \includegraphics[height=.95\textwidth]{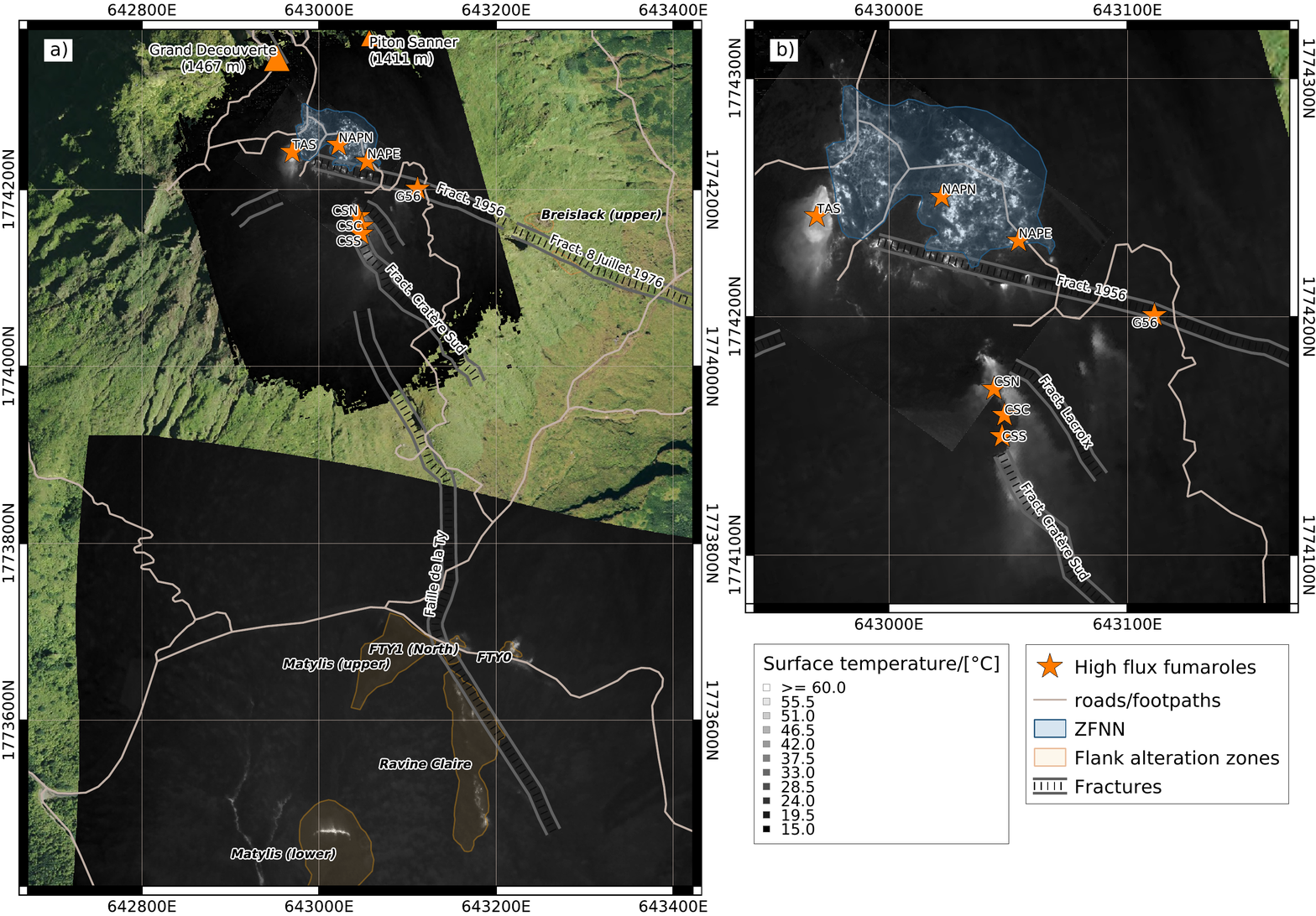}
  }
  \caption{\added{a)} Georeferenced thermal images of La Soufrière 
    \deleted{de Guadeloupe} and
    surroundings taken from the helicopter on 22 \replaced{November }{Nov.~}
    2019 between 05:40 and
    06:05 local time \deleted{at a) 1:5000} b) \replaced{zoom}{and at
      1:2000 scale} showing the summit thermal anomalies.
    The base map is the 2017 IGN aerial orthophoto (BDOrtho).  The
    thermal images are shown in greyscale where white and black denote hot
    and cold, respectively.}
  \label{fig:georeferenced_thermal}       
\end{figure*}

\begin{figure*}[ht]
  \centering
  \includegraphics{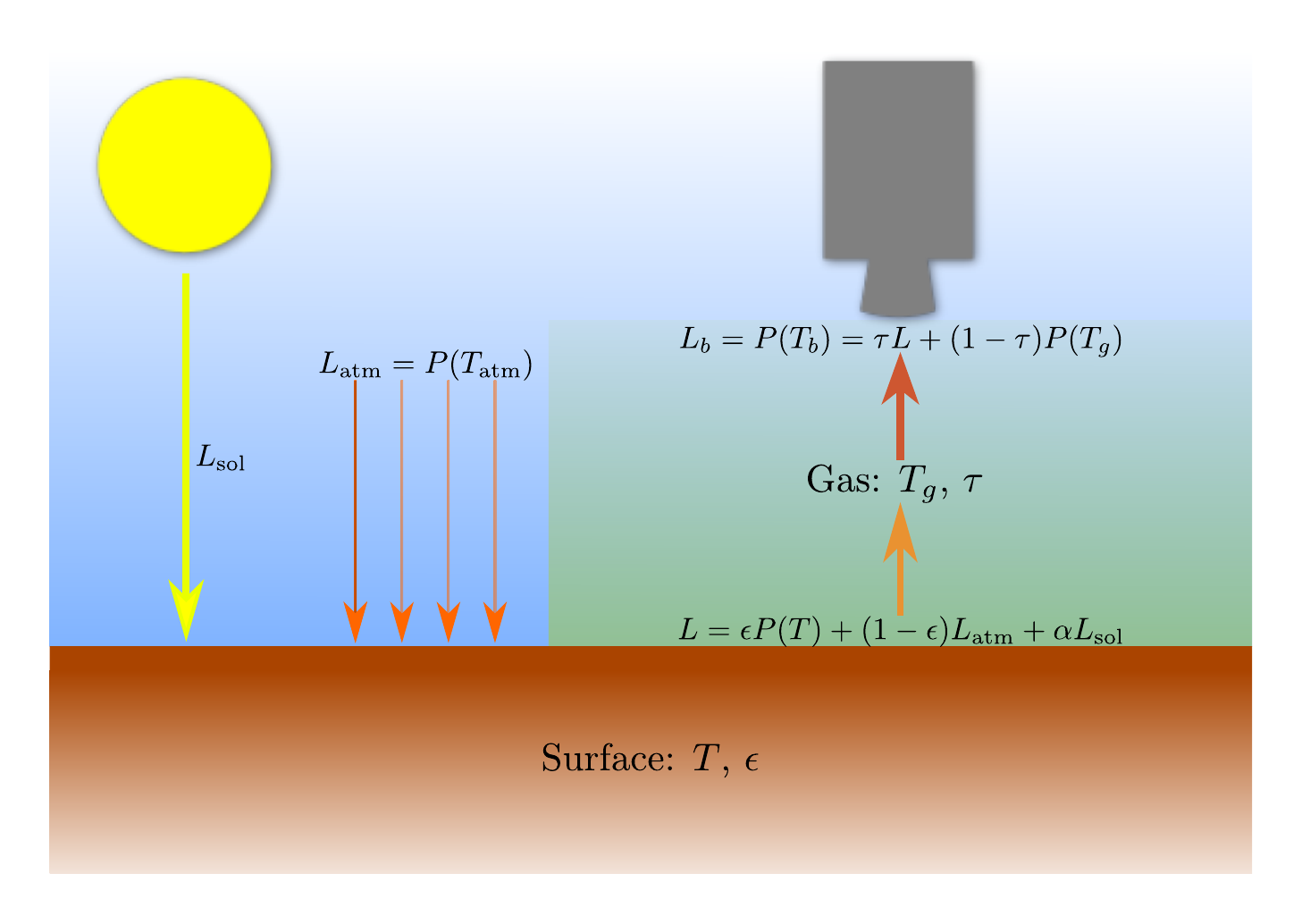}
  \caption{Conceptual model of the heat flux measured by a thermal camera,
    \added{$L_b$} viewing  and displayed as the ``brightness'' temperature,
    \added{$T_b$}.  The incoming heat fluxes (left) from solar radiation,
    $L_\text{sol}$, and from the
    atmosphere, $L_\text{atm}$, are reflected from the surface in proportion
    to the surface albedo, $\alpha$, and $1-\epsilon$, respectively, where
    $\epsilon$ is the emissivity of the surface.  For a long-wave infrared
    sensor such as the thermal camera used here, $\alpha\approx 1-\epsilon$.
    The emitted radiation of the surface\replaced{, $L$ (which depends on 
      surface temperature, $T$, though the Planck function, $P$)}{ (a
      function of the surface temperature, $T$)}, is added to these
    reflected fluxes which arrive at the
    camera having been transmitted through a mixture of atmospheric and
    volcanogenic gases at temperature, $T_g$, and having transmissivity,
    $\tau$.}
  \label{fig:stock_flow}       
\end{figure*}

We show our results from the analysis of the thermal images (Fig.\:\ref{fig:georeferenced_thermal}) in Table \ref{tab:Table1}.  For each site with detected thermal activity (summit, lower Ravine Matylis, Ravine Claire and FTY), we have determined radiative and convective fluxes as well as the flux density, $q_i = Q_i / A$.  As large fluxes can be observed by low intensity emissions over a large area, we also calculate the total heat flux density, $q = (Q_{\mathrm{rad}} + Q_{\mathrm{conv}}) / A_{heated}$, as a metric for comparing intensity between sites (Table \ref{tab:Table1}) with a relative standard error of about 6\%.  
At the summit we found the radiative flux to be \SI{0.74}{\MW} and the convective flux to be \SI{4.94}{\MW}, with $h_c = $\:\SI{41.6(23)}{\Wmsq\per\kelvin} (Eq. \ref{eq:schlichting_neri}) and $A_{\mathrm{heated}} =$\:\replaced{\num{14070}}{\num{14074}}\,\si{\msq} \added{(areas were calculated by counting the number of heated pixels, e.g.\:Fig.\:\ref{fig:georeferenced_thermal})}.  These values were by far the largest in magnitude of all the sites, and larger than the total heat fluxes for the other sites combined.  This finding is supported by the heat flux density, which is considerably greater than any other site.

We calculated $h_c = $\:\SI{37.8(23)}{\Wmsq\per\kelvin} for wind speeds during acquisition of the images of the flank sites, which was used in the calculations for all sites.  Owing to a relatively large emitting surface of \replaced{\num{8010}}{\num{8014}}\,\si{\msq}, the heat flux at Ravine Claire (RC) is second to the summit with radiative and convective fluxes of \SI{0.13}{\MW} and \SI{0.75}{\MW} (Table \ref{tab:Table1}), respectively.  In the lower Matylis ravine, a strong thermal anomaly leads to high flux densities (\replaced{42}{41.7} and \SI{238}{\Wmsq} for radiation and convection, respectively), although a low heated area (\SI{1630}{\msq}) keeps the overall fluxes low.  We identified two sites along FTY \added{(FTY0 and FTY1 in Figs.~\ref{fig:soufriere_overview} and \ref{fig:georeferenced_thermal})} which have similar results for flux density and had a total flux of \SI{0.58}{\MW} and a mean flux density of \replaced{\num{227}}{\num{226.7}}\,\si{\Wmsq}.  We note that all these sites (Matylis, RC, FTY) are linked to the Ty N-SE and Galion N-S faults that cut the dome \citep{KOM:Gwada2005, ROS:SciRep2016}.

\subsection*{Fumarole heat and mass flux}

\begin{figure*}[ht]
  \includegraphics[width=\textwidth]{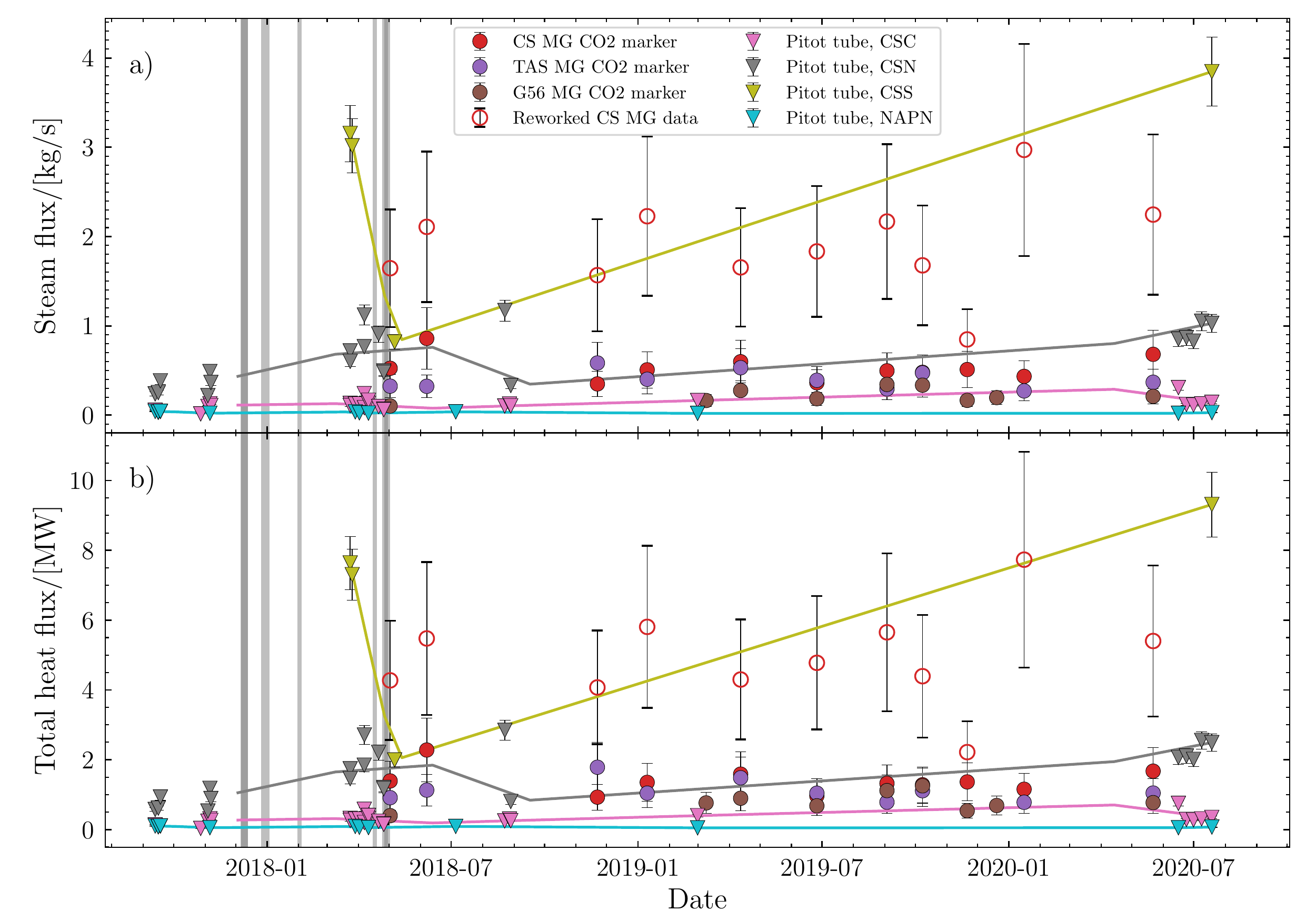}
  \caption{Time series of the summit fumarole fluxes since the last quarter 
    of 2017 as estimated from pitot tube and multigas data.  Steam fluxes are
    shown in a) and heat fluxes in b).  Vertical grey bars indicate the 
    record of VT earthquakes with magnitude $> \mathrm{M}2.0$ (including
    felt VTs), which are 
    limited to a sequence in early 2018.  To aid interpretation of the pitot 
    tube data, we used a linear smoothing function to produce the dashed 
    curves.  The MultiGAS data shown (filled symbols) incorporate the
    ~35\% increase
    in steam flux due to condensation of vapour within the plume.  The
    reworked MultiGAS data correspond to the CO$_2$ flux multiplied by the
    H$_2$O/CO$_2$ ratio determined from Giggenbach bottle analyses.}
  \label{fig:short_time_series}       
\end{figure*}

The mass and heat fluxes are shown in Fig.\:\ref{fig:short_time_series} a) and b), respectively.  Steam fluxes estimated from MultiGAS traverses show that, using CO$_2$ as a marker, the min/mean/max values are: 0.35/0.52/0.86, 0.15/0.30/0.47 and 0.29/0.44/\SI{0.67}{\kg\per\second} for CS, G56 and TAS, respectively.  Heat flux estimates based on these data give 0.93/1.36/2.28, 0.40/0.79/1.25 and 0.78/1.12/\SI{1.79}{\MW} for CS, G56 and TAS, respectively.  Considering the relative standard error of 40\%, we find that the MultiGAS fluxes have remained stable since regular estimates began in mid 2018, subsequent to the M4.1 earthquake.

The temporal variations in fumarole steam flux calculated by the Pitot-tube for the three vents at CS are also shown in Fig.\:\ref{fig:short_time_series}a.  These data indicate that the fluxes can show a large degree of variation in short time periods which is especially true during periods of accelerated unrest such as from March\replaced{--}{-}May 2018 \citep{MOR:JVGR2020}.  We find steam fluxes to have min/mean/max values of 0.01/0.12/\SI{0.31}{\kg\per\second} at CSC, 0.22/0.70/\SI{1.50}{\kg\per\second} at CSN and 0.82/2.71/\SI{3.85}{\kgs} at CSS (Fig.\:\ref{fig:short_time_series}a), which equate to heat fluxes of 0.03/0.29/\SI{0.75}{\MW} for CSC vent, 0.53/1.69/\SI{3.64}{\MW} for CSN vent, and 1.99/6.56/\SI{9.31}{\MW} for CSS vent (Fig.\:\ref{fig:short_time_series}b).  Fluxes were also measured at NAPN and we found mean mass and heat fluxes of \SI{0.03}{\kgs} and \SI{0.07}{\MW} with very little variation over time, including during the 2018 unrest.  The contribution of NAPN to the total heat and mass budget is thus negligible.  We note that, due to an improved method for estimating vent area based on head-on thermal images compared to visual estimation during measurements (see Methods), the vent heat fluxes presented here are quantitatively lower than reported in \cite{MOR:JVGR2020}, although the qualitative temporal variation is the same.  The Pitot-tube data show that vent fluxes at CS were strongly affected by and decreased during the 2018 unrest phase, but have since settled to around \SI{4}{\kgs} and \SI{10}{\MW} for mass and heat flux, respectively.  Fig.\:\ref{fig:short_time_series}b) shows, although the coefficients for heat capacity and latent heat vary with temperature, the same trends as per Fig.\:\ref{fig:short_time_series}a), indicating that the heat flux depended much more strongly on the variations in mass flux than temperature changes during this period.

At CS, we have overlap in the Pitot-tube and MultiGAS instrument data that allows us to compare the data collected by these instruments from closely spaced outings.  For example, in terms of steam flux the Pitot-tube data from \replaced{15 June 2020}{2020-06-15} show that CSC+CSN+CSS emitted around \SI{4.6(5)}{\kgs}.  The flux estimated from MultiGAS measurements at CS on \replaced{22 May 2020}{2020-05-22} were \SI{0.7(3)}{\kgs}.  These MultiGAS estimates are almost an order of magnitude times lower than those from the Pitot-tube, and Fig.\:\ref{fig:short_time_series} indicates that this is systematically the case.  Whilst we have attempted to correct for the quantity of condensed vapour that is undetectable by the MultiGAS, additional errors in this calculation are likely to be primarily responsible for the difference between these two values, although they agree to within an order of magnitude and appear to show qualitatively the same temporal variations.

\subsection*{Thermal springs heat and mass fluxes}

\begin{figure*}[ht]
  \includegraphics[width=\textwidth]{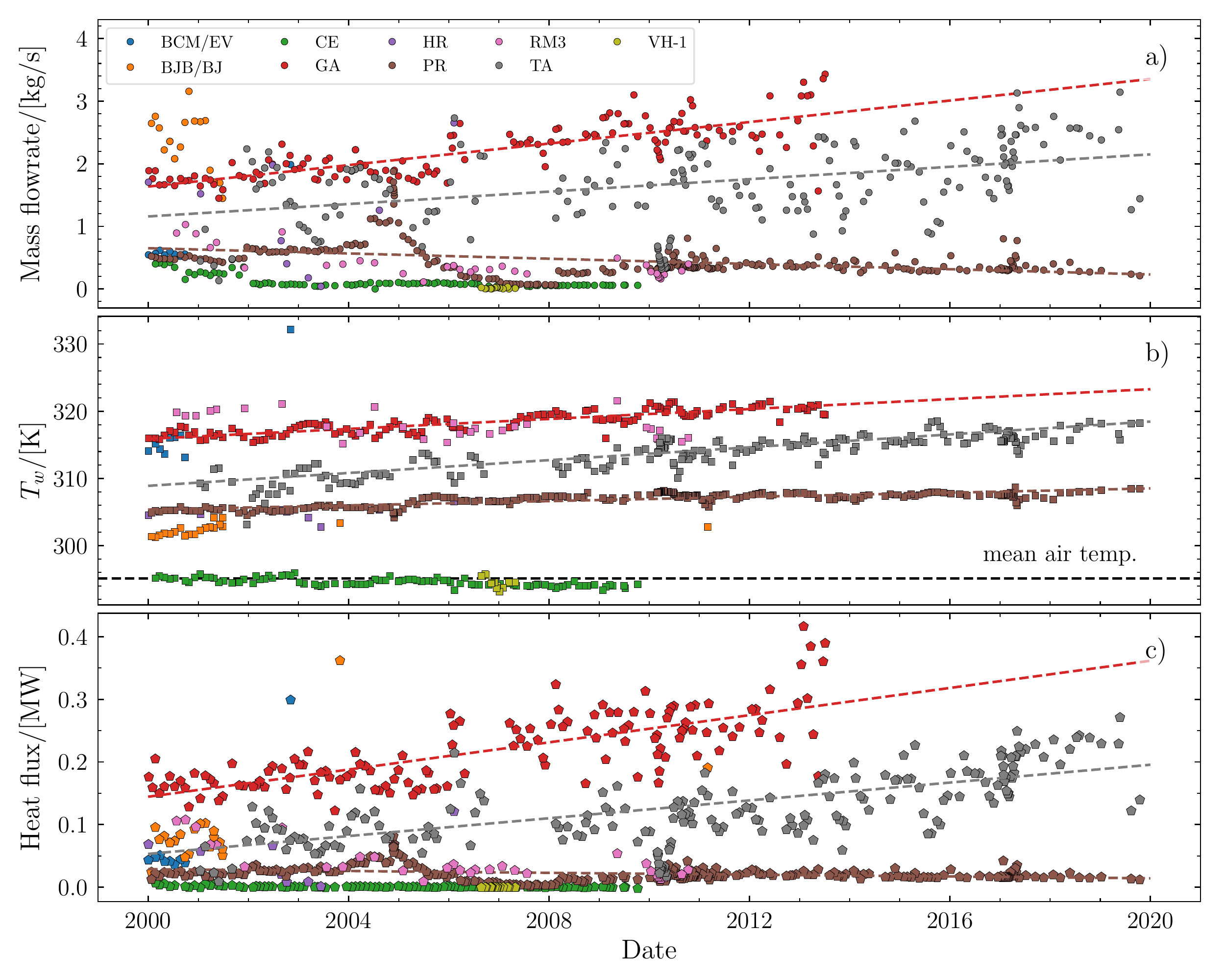}
  \caption{Mass flow rate, water temperature and heat flux for the thermal
    springs monitored by the OVSG for the period from 2000-2020.  See text 
    for site codes.  The colour code for each site is identical between plots
    and dashed lines show linear trends for the GA, TA and PR sites.  }
\label{fig:thermal_springs}       
\end{figure*}

In Fig.\:\ref{fig:thermal_springs}, we present the mass flow rate and temperature measured over the period between 2000 and 2020 \citep[a subset of the entire data set, see][Fig.\:\ref{fig:thermal_springs}a and b]{VIL:EPSL2005}, as well as the heat flux calculated from this via Eq.\:(\ref{eq:spring_fluxes}) (Fig.\:\ref{fig:thermal_springs}c).  Whilst the flow rate and temperature measurements have continued until the present day, there are gaps in the mass and heat flux data during 2014\replaced{--}{-}2016 due to instrument failures.  The GA, Tarade spring (TA), Bains Jaunes (BJ) and Pas du Roy (PR) springs are amongst the most accessible and this is reflected in both the abundance and persistence of the measurements in the OVSG database.  They are also the most representative of acid-sulphate thermal springs linked to La Soufrière\deleted{ de Guadeloupe}'s hydrothermal activity.  This record does not reflect the absolute total mass/thermal output of the thermal springs as i) other sites are known but are far \replaced{less accessible}{more inaccessible} or impractical to measure and ii) some sites may not yet have been discovered.  However, particularly as GA and TA have the largest known flow rates, it is likely that these calculations are nonetheless representative of the total budget for the thermal springs.  We fitted linear trends to the data for TA, GA and PR, and extrapolated where necessary to project the values to the current date.

Overall, we see that both mass flow rate and water temperature have slowly and steadily increased over time in an approximately linear fashion.  For example, the flow rate at TA increased from around \SI{1.1}{\kgs} in 2010 to \SI{2.1}{\kgs} at present whilst its temperature rose from 309 to \SI{318}{\kelvin} \deleted{(36 to 45\,$^\circ$C)}.  Only the TA and PR sites have data that cover the whole data range and manual measurements stopped at GA in 2014.  Historically, GA dominates the heat budget for the thermal springs, and has almost double the output of TA.  Summing over these three sites, we find that the total heat flux from the thermal springs is around \SI{0.57}{\MW} \added{(including the extrapolated trend for GA)}.

%
\section*{Discussion}

\subsection*{Comparison of steam and heat flux estimation methods}

\subsubsection*{Fumarole flux}

Our measurements (Fig.\:\ref{fig:long_time_series}) show that the plume mass and heat fluxes have not undergone extensive evolution since August 2005.  With this in mind, we must consider that the fumarole plume heat and mass flux estimates of \cite{GAU:JVGR2016} to be excessively high.  In their discussion, the order of magnitude discrepancy with the estimations from MultiGAS traverses \citep{ALL:CG2014} was mostly attributed to the MultiGAS studies not accounting for condensed water vapour.  However, we note several key assumptions in \cite{GAU:JVGR2016} that may have led to systematic errors in their estimations:
\begin{itemize}
  \item Plume thickness grows \added{linearly} with distance from the vent
    \added{, $x$,} \deleted{as $\mathcal{H} \sim x$,} and not $x^{2/3}$.
    \replaced{This latter scaling is for}{which is the scaling of} the 
    height of the plume axis \added{from the ground} \citep{SLA:JFM1967},
    so that their mass flux integral overestimated the plume area.
  \item 
    \added{The plume section was assumed to be axisymmetric though this is
    generally not true: wind-blown plumes from smoke stacks, cooling towers
    and in laboratory experiments have been shown to be more broad
    (horizontally) than thick \citep[vertically,][]{CON:AE2001}.  Thus,
    transmissivity calculated looking horizontally through a plume would have
    been lower than was actually the case, leading to an overestimation of
    plume temperature.  Consequently the plume density and thus the mass flow
    rate should be lower than the given estimations.}
    \deleted{In calculating plume transmissivity, the plume section is 
    assumed to be axisymmetric.  However, this is generally not correct: 
    wind-blown plumes from smoke stacks, cooling towers and in laboratory 
    experiments have been shown to be more broad (horizontally) than thick
    Thus, calculated plume
    transmissivity looking horizontally through a plume will be lower than
    is actually the case and plume temperatures will be higher.
    Consequently the plume density and thus the mass flow rate will be lower
    than they estimated.}
  \item The vapour carrying capacity of the plume was assumed to be equal to
    that of the atmosphere.  However, as plume temperatures are higher than
    atmospheric and more water vapour can therefore be carried without
    condensing\replaced{,}{ so} this relationship does not hold.
\end{itemize}

Overall, this suggests that a more realistic plume flux for 2010 would be more in line with the MultiGAS (taking into account condensed vapour) and Pitot-tube measurements, that is a steam flux of \SI{5.3}{\kgs} for CS.  \replaced{Thus taking the Pitot-tube measurement at CS as the ground truth, we found a scaling factor for the MultiGAS measurements.  Using this to scale the MultiGAS flux for TAS gives \SI{6.2}{\kgs}.  Likewise, we find heat fluxes of \SI{13.0}{MW} for CS and \SI{15.2}{\MW} for TAS (see Table \ref{tab:Table2}) for the 2010 study.}{Scaling the flux accordingly for TAS gives 6.2 kg/s.  In terms of heat flux, we thus find 13.0 MW for CS and 15.2 MW for TAS (see Table \ref{tab:Table2}).}  

Moreover, despite some assumptions, our results show that MultiGAS traverses and Pitot-tube measurements provide qualitatively coherent vent flux estimations yet quantifying the steam flux using MultiGAS is a challenge, particularly in a tropical atmosphere with 100\% RH.  To overcome this shortcoming, we take the H$_2$O/CO$_2$ ratio determined from Giggenbach bottle sampling \citep[OVSG-IPGP 2017-2020;][]{MOR:IJG2020} and multiply this by the CO$_2$ flux estimated from the MultiGAS data.  As this ratio is measured at the vent, it is not subject to a loss of matter due to condensation contrary to measurements within the plume.  The resulting fluxes at CS resemble much more closely the Pitot-tube-derived fluxes (see ``Reworked CS MG data'' in Fig.\:\ref{fig:short_time_series}).  This correlation starkly indicates the difficulties in accounting for condensed volcanogenic vapour in the MultiGAS steam-flux estimations.  Nevertheless, in a monitoring context, either or both methods could be applied in various volcanoes worldwide to estimate their mass and heat fluxes.

\subsubsection*{Ground flux}

Although we have used the same model for $h_c$ as \cite{GAU:JVGR2016}, we obtain slightly different values simply due to differing weather conditions (compare $h_c = $\:\num{41.6(23)} and \SI{37.8(23)}{\Wmsq\per\kelvin} for the summit and flanks, respectively, with $h_c = $\,\SI{30.5(3)}{\Wmsq\per\kelvin} as derived from data in Table \ref{tab:Table2} of \citealp{GAU:JVGR2013}\added{)}.  Thus, similarity between the results of our study and those of \cite{GAU:JVGR2013} at the same site would be suggestive of a decrease in temperature at that site.  A good comparison can be made at the FTY sites.  We note in particular that the mean total heat flux density for FTY0 + FTY1 sites combined, \SI{228(14)}{\watt\per\msq}, is in strong agreement with the heat flux density calculated from temperature gradient measurements of \SI{265(45)}{\Wmsq} \citep{GAU:JVGR2013}, which suggests that, on average, temperatures have not decreased (the ambient temperature during the 2010 survey and ours was approximately \SI{17}{\celsius} in both cases).  It is somewhat unclear precisely where \cite{GAU:JVGR2016} defined the boundary of FTY and, indeed, their Fig.\:\ref{fig:soufriere_overview} suggests that this might extend into what we define as Ravine Claire, so judging the evolution of the extent of this site is difficult.  However, taking uncertainties into account, the present-day total flux for FTY0+FTY1+RC of \SI{1.46(23)}{\MW} is not too dissimilar to the \SI{1.0(2)}{\MW} reported in 2010.

\deleted{As stated above, images were taken from about 100--300\,m above the ground.  }Given the angular resolution of the thermal camera\added{ and camera-to-ground distances of 50--\SI{300}{\metre}, this equates to a pixel length of between 0.08 and \SI{0.5}{\metre}}.  As fumaroles are often of smaller dimension than the resolution length-scale, especially when recording from greater distances, their temperature is integrated over the pixel area along with the cooler surroundings and so pixels that cover thermally non-homogeneous ground will display temperatures lower than the true temperature of the hotter object.  \cite{HAR:BV2009} demonstrated that, at a distance of 100 m, the pixel-integrated temperature of a $6 \times 13$ cm (\SI{78}{\centi\msq}) fumarole in a \SI{169}{\centi\msq} pixel was lower than the actual temperature by \SI{40}{\celsius}.  Taking this into account, we must consider that the fluxes that we calculate are minimum estimates, emphasising the importance of the ground heat flux for La Soufrière\deleted{ de Guadeloupe}.

\subsection*{Total heat budget}

As noted by \cite{GAU:JVGR2016}, some heat loss may be undetectable by the methods described in this work, due to either vegetation cover (e.g. on the flanks) or temperature changes that are below the instrument resolution (cf. summit in the region of CS).  As per their work, we note that the ``background'' heat conducted through the system, as deduced from borehole measurements and extrapolated to the scale of the dome adds only an additional \SI{0.013}{\MW}.  Furthermore, we note that some heat will be transported \replaced{by gases other than steam}{not by steam but by other gases}, notably CO$_2$ in the plume and, in particular, CO$_2$ soil degassing which is a widely-used proxy for heat flux \citep[cf.][]{CHI:ApplGeochem1998, BLO:GGG2014, HAR:JVGR2015}.  A detailed study is beyond the scope of this present work but we may make progress under the following assumptions:
\begin{enumerate}[1.]
  \item Passive CO$_2$ degassing occurs in the same areas and to the same extent as the ground thermal anomalies.
  \item The ground heat flux, $Q_{\mathrm{soil}}$, calculated above equals the underground convection of steam, $\dot{m}_{H2O} c_{p,H2O} (T - T_{\mathrm{amb}})$.
  \item The CO$_2$/H2O ratio in areas of soil degassing is the same as in the fumaroles \added{\citep{CHI:JGR2001}}.
\end{enumerate}

Under assumption (2), $\dot{m}_{H2O} = $\:\SI{135.5}{\kgs} based on a typical anomaly temperature of \SI{80}{\celsius} and ambient temperature of \SI{15}{\celsius}.  Based on a H$_2$O/CO$_2$ ratio of 43.5 (OVSG-IPGP, 2020), this gives $\dot{m}_{CO2} = $\:\SI{3.2}{\kgs} and thus $Q_{\mathrm{CO2}} = \dot{m}_{\mathrm{CO2}}\,c_{\mathrm{p,CO2}} (T - T_{\mathrm{amb}}) = $\:\SI{0.19}{\MW}.

Clearly, although comparable to the contributions of certain thermal springs, heat transport by CO$_2$ does not add significantly to the total budget.  Nevertheless, due to the accelerating spread of the altered zones and the fact that the area over which CO$_2$ degassing occurs may be far greater than that involved in degassing of water vapour \citep[cf.][]{CHI:JGR2005}, the OVSG has begun to carry out joint surveys of soil temperature profiling and CO$_2$ flux \citep[via the accumulation chamber technique,][]{CHI:ApplGeochem1998} in order to further constrain ground heat losses and we will return to this issue in a forthcoming paper.

Summing the fumarole (\SI{28.3(68)}{\MW}, 77.5\%), ground (\SI{7.6(11)}{\MW}, 20.8\%\added{)} and the thermal springs fluxes (\SI{0.56}{\MW} 1.5\%), we obtain a total heat output of \SI{36.5(79)}{\MW} (see Table \ref{tab:Table2}).

\subsection*{Temporal evolution}

La Soufrière\deleted{ de Guadeloupe} has undergone a significant evolution of its activity \replaced{during 2010--2020}{in the past decade} as described and analysed in detail by \cite{MOR:JVGR2020} and OVSG-IPGP (2014-2020).  This can be summarised as follows:
\begin{enumerate}[1.]
  \item The appearance of new fumarolic vents and the reactivation of pre-existing fumaroles with local\deleted{ly} high-velocity degassing.
  \item Vegetation die-off near-to and far-from active vents (see Supplementary Material Fig.\:1).
  \item The enlargement of a major extensive area of heated ground on the summit areas that progresses towards the North from the Fracture Napoléon (see below and Supplementary Material Fig.\:2).
  \item More frequent and stronger seismic events including felt events (M4.1, April 2018).
  \item An acceleration in the opening rate of several summit fractures.
  \item The appearance of new mineralised water springs at the base of the volcano as a result of the rapid cooling of hydrothermal fluids.
\end{enumerate}

Undoubtedly the greatest phenomenological change at the summit of La Soufrière\deleted{ de Guadeloupe} is indeed the appearance and spread of the ground thermal anomaly in the ZFNN region.  For example, the heated area at the summit has gone from an estimated \SI{610}{\msq} in 2010 \citep{GAU:JVGR2016} to \SI{14070}{\msq} for the present study.  Indeed, whereas \cite{GAU:JVGR2016} identified thermal anomalies along the Napoléon fracture and in the craters containing the CS fumaroles, they calculated that the associated heat losses were \SI{0.01}{\MW} by radiation and \SI{0.2}{\MW} by forced convection.  The present-day \replaced{radiative and convective fluxes of \num{0.74(7)} and \SI{4.94(49)}{\MW}}{convective and radiative fluxes of $4.94 \pm 0.49$ MW and $0.74\pm0.07$ MW} have increased by an order of magnitude \replaced{between 2010 and 2020,}{in the past decade} which is in large part due to this increase in heated area and also, to a lesser degree, because of increased temperatures.  The total heat flux density presently estimated at \SI{403(26)}{\Wmsq} is greater than the 2010 estimate of \SI{326.3(685)}{\Wmsq}, and thus suggests an increase in thermal intensity at the summit, though these values are within the bounds of measurement uncertainty.

\begin{figure*}[ht]
  \includegraphics[width=\textwidth]{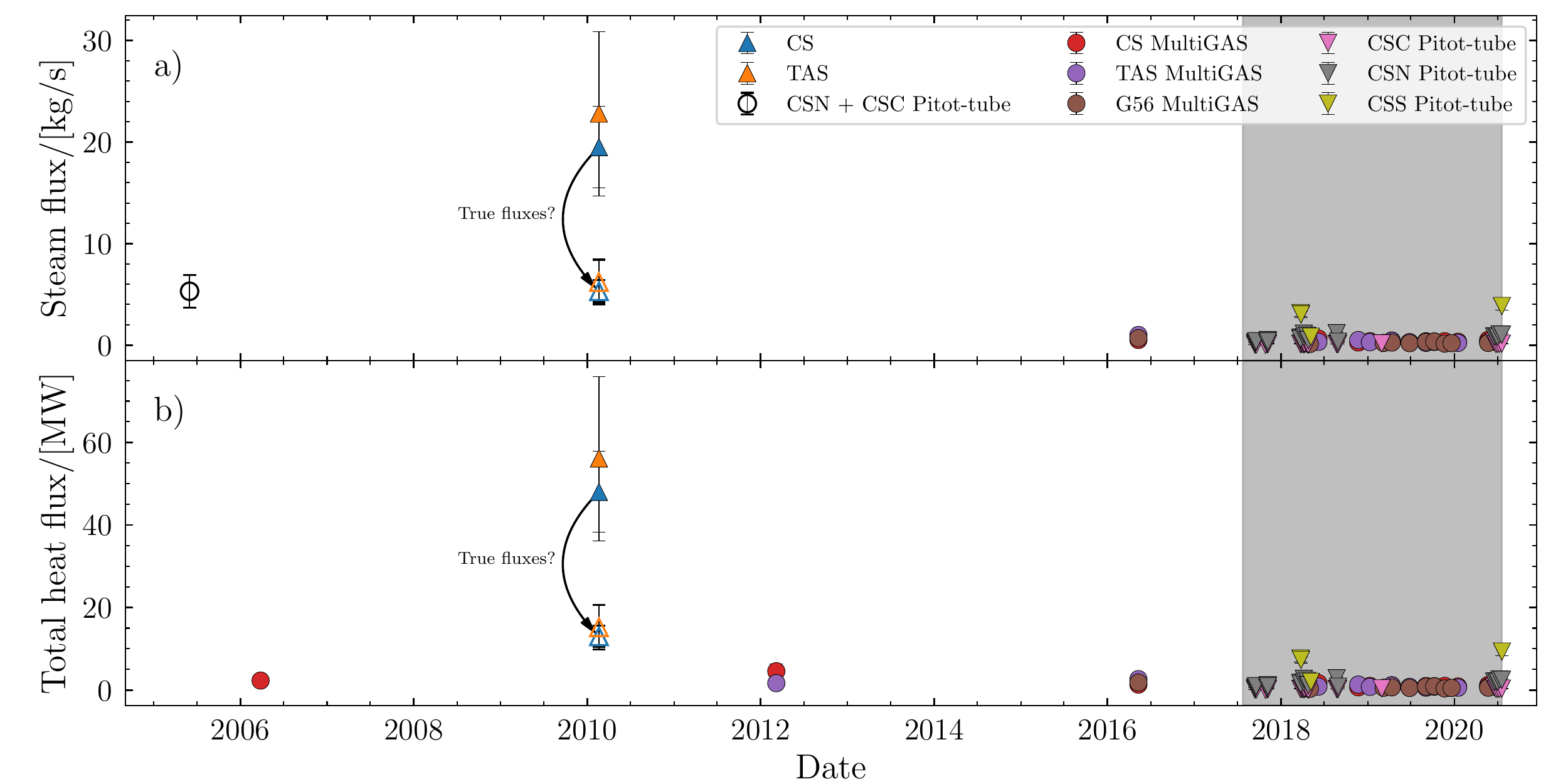}
  \caption{Long term temporal evolution of the fumarole fluxes from 
    2005-present.  In addition to the present dataset, we also plot 
    Pitot-tube data from March-May 2018 \citep{MOR:JVGR2020} and 2005, along 
    with thermal camera data 2010 (both \cite{GAU:JVGR2016}, MultiGAS data
    from 2006, 2012 \citep{ALL:CG2014}, and reworked data from 2016-2017
    \citep{TAM:Geosci2019}.
    The arrow labelled ``True fluxes?'' indicates our reworking of the 2010
    thermal camera data.  
    Grey shading indicates the period covered by the
    present dataset, as shown in Fig.\:\ref{fig:short_time_series}.}
  \label{fig:long_time_series}       
\end{figure*}

Apart from the pulse of unrest around March\replaced{--}{-}April 2018, the fumarolic fluxes have not changed considerably since early 2018 (Fig.\:\ref{fig:short_time_series}), and the thermal spring fluxes have increased only slightly (Fig.\:\ref{fig:thermal_springs}) over the past 20 years.  To gain a greater perspective of the overall temporal evolution of plume fluxes over a similar period, we plot in Fig.\:\ref{fig:long_time_series} our data along with the steam and heat fluxes taken from \cite{ALL:CG2014}, \cite{GAU:JVGR2016} and \cite{TAM:Geosci2019}.  This figure shows that our current data are, given the natural variability of these fluxes, consistent with the previous MultiGAS measurements of \cite{ALL:CG2014} and \cite{TAM:Geosci2019}\replaced{.  They are far more consistent with}{ and also} the Pitot-tube measurement for the CSN + CSC vents cited in \cite{GAU:JVGR2016} \added{than the fluxes that they derived from analysis of thermal images} (compare their value of \SI{5.3(16)}{\kgs} to the contemporary sum for CSN + CSC of around \SI{1.4(3)}{\kgs}, Fig.\:\ref{fig:short_time_series}a).

Three major swarms of VT earthquakes occurred from 1 February to 28 April 2018, with the third swarm initiated by the off-volcanic axis M4.1 earthquake which struck at 00:32 UTC on 28 April and was widely felt throughout Guadeloupe.  In particular, as reported in \cite{MOR:JVGR2020}, a short-lived increase in plume flux occurred concurrently with temperature increases before the earthquakes, but both observables had returned to background levels before the M4.1 event on 27 April 2018.  Hence, our Pitot-tube flux results illustrate the importance of these flux estimations for close monitoring of volcanic activity.

Our results combined with published data indicate that plume flux has decreased overall since these records began.  Thus, given a lack of increased VT seismicity or other signs of sudden evolution in 2010, and in the light of the errors discussed above, it seems that the values reported in \cite{GAU:JVGR2016} are anomalous.  In order to provide a better comparison with the present study, we attempt to re-estimate the 2010 plume fluxes given the available data from this period \citep{ALL:CG2014, GAU:JVGR2016}.  We suppose that, despite possible overestimation, the ratio of CS/TAS fluxes was correctly established in 2010 and that the relative standard error will remain unchanged.  Thus scaling with the 2005 Pitot-tube data, for 2010 we find steam fluxes of \SI{5.3(11)}{\kgs} for CS and \SI{6.2(22)}{\kgs} for TAS, and heat fluxes of \SI{13.0(26)}{\MW} for CS and \SI{15.2(54)}{\MW} for TAS (Table \ref{tab:Table2}).

We recalculate the thermal spring fluxes for 2010 based on an interpolation of the thermal springs time series the date of the aerial survey (see Fig.\:\ref{fig:thermal_springs}).
These new values for the thermal springs fluxes are, in general, not appreciably different to the present values, except for TA which we find to be about half the reported value for 2010.  Additionally, this process does allow us to calculate fluxes for PR, absent in the 2010 survey.  Consequently, we find that the total heat flux in 2010 was likely to have been around \SI{30}{\MW} of which the fumarolic contribution was approximately \SI{28}{\MW}, or 95\% of this total (see Table \ref{tab:Table2}).  This compares to the estimate of 98\% by \cite{GAU:JVGR2016}.

The starkest change in the past decade is an increase in ground heat flux by greater than an order of magnitude (see Table \ref{tab:Table2}), reflecting the appearance of the strong and widespread thermal anomaly in the ZFNN area in particular.  Indeed, the ground heat flux has increased from \SI{1.2(3)}{\MW} of a total budget of \SI{29.8(83)}{\MW} (4\%) to \SI{7.6(11)}{\MW} of \SI{36.5(79)}{\MW} (21\%).  Nevertheless, in terms of heat flux density across the entire edifice, the present-day mean value of \replaced{\SI{403(24)}{\Wmsq} is slightly higher than the \SI{337.1(1966)}{\Wmsq} estimated in 2010}{$288.8 \pm 88.9$ W/m2 is very similar to $337.1 \pm 196.6$ W/m2 in 2010}.
\replaced{Using the reworked values from}{Going by the values published in} \cite{GAU:JVGR2016}, fumarole heat flux has \replaced{decreased at CS and TAS}{quite drastically decreased}, \replaced{decreasing}{going} from \added{a total of} \SI{28.2(80)}{\MW} to \SI{17.8(45)}{\MW} (\replaced{see}{excluding G56, } Table \ref{tab:Table2} and Fig.\:\ref{fig:long_time_series}).
\added{However, if we also include G56, our reanalysis shows that the overall change in output is insignificant and, along with the appearance of NAPN and NPE fumaroles (which contribute negligibly to the total budget), highlights a spreading of fumarole output over the dome.}
\deleted{However, our reanalysis suggests that this change is not so important ($28.2 \pm 8.0$ MW, Table \ref{tab:Table2} and Fig. \ref{fig:long_time_series}).}
Heat transport at the thermal spring sites has increased from 0.4 to \SI{0.6}{\MW}.
Taken together, these findings are suggestive of the edifice becoming more fractured and porous over time, allowing some of the hydrothermal fluids to escape via different pathways.  The increase in porosity and fracturing may be a result of rock dissolution and weakening \citep[e.g.][]{POLA:Tecto2012, WYE:JVGR2014, HEAP:JVGR2015, MOR:BV2019} as a result of hydrothermal alteration\added{, mainly} by acid-sulphate fluids \citep{SAL:JGE2011}.  An increase in porosity and fractures is coherent with the increased opening rate of fractures as well as the GNSS radial horizontal displacements of 3--10\,mm/year \citep{MOR:JVGR2020} and is corroborated by the appearance of mineralised deposits at the base of the dome in the upper northern branch of the Matylis river and in the Breislack area (Fig.\:\ref{fig:soufriere_overview}b).

\added{Given recent unrest events, these results are coherent with recent petrological analysis of the volcano's last major magmatic eruption which place a shallow magma reservoir at 5--\SI{8}{\km} depth \citep{PIC:JP2018}.
  Furthermore, the interpretation of the geochemistry of emitted fluids \citep{VIL:JVGR2014, MOR:JVGR2020, MOR:IJG2020} infers a contribution of deep magmatic gases that trigger periodic transitory heating and pressurisation of the deep hydrothermal system near the critical point of water at 2--\SI{3}{\km} below the summit.
  Despite a lack of further evidence for the ascent of magma to very shallow depths, it is important to keep in mind from a hazard perspective, that larger, more frequent, or more intense transitory pulses of hot magmatic gases could exceed the buffering capacity of the hydrothermal system bringing the overall system to critical conditions compatible with phreatic/hydrothermal eruptive unrest.}
Consequently we must expect that the extent and magnitude of thermal anomalies and diffuse degassing on La Soufrière\deleted{ de Guadeloupe} will continue to increase over time and, indeed, this is evident in aerial images (cf.\:Figs.\:A.1 and A.3, supplementary material).

\subsection*{Comparison with other hydrothermal systems}

La Soufrière\deleted{ de Guadeloupe}'s total heat budget is on par with other major hydrothermal system volcanoes.  For example, the heat output at Vulcano (Italy) was estimated at 10--\SI{12}{\MW} from combining ground based radiometer and ASTER measurements \citep{MAN:GRL2019}.  At Whakaari\deleted{ (NZ)}, heat output estimated using crater floor soil CO$_2$ degassing as a proxy was found to be \SI{20.0(25)}{\MW} \citep{BLO:GGG2014}.  Our value is somewhat smaller than Nisyros (Greece), Ischia or Campi Flegrei (both Italy) which release in the range of 40--\SI{100}{\MW} \citep{CHI:JGR2005}.  However, as noted by Harvey et al. (2015), whilst total heat budget is helpful for following the temporal evolution of an individual volcano, a more useful metric for comparing between systems is the heat flux density as, in many of these cases, the major component of heat flux is through diffuse soil heating so larger systems naturally tend to emit more heat.  The mean ground heat flux density (combining radiative and convective fluxes) for the entire La Soufrière\deleted{ de Guadeloupe} complex is \SI{406(24)}{\Wmsq} (Table \ref{tab:Table1}) which, if we consider the total heat budget over the total heated area of \SI{26280}{\msq}, the mean flux density of the currently active part of the La Soufrière complex climbs to \SI{1366(82)}{\Wmsq} (Tables \ref{tab:Table1} and \ref{tab:Table2}).  Based on the data compiled in \cite{HAR:JVGR2015} from heat flux density calculated from CO$_2$ flux, this ranks La Soufrière\deleted{ de Guadeloupe} amongst the world's most powerful heat producing volcanoes, well above Whakaari (\deleted{NZ - }205 W/m2), Vulcano \citep[\SI{140}{\Wmsq}, ][]{MAN:GRL2019}, Campi Flegrei (\SI{118}{\Wmsq}\added{)} and Nisyros (\SI{19}{\Wmsq}), and roughly on par with Ischia (\SI{764}{\Wmsq}).  Similar to La Soufrière\deleted{ de Guadeloupe}, Vulcano, Whakaari and Ischia are also dome volcanoes and the larger heat flux densities here may indicate optimal steam transport to the surface along high-permeable pathways associated with dome emplacement: Ischia, in particular, has a fumarolic H$_2$O/CO$_2$ ratio similar to that at la Soufrière\deleted{ de Guadeloupe}, \citep[H$_2$O/CO$_2$=147 in 2001,][]{CHI:JGR2005}.
  This, due to the very extensive hydrothermal system at La Soufrière\deleted{ de Guadeloupe}, indicates the dominance of heat and mass transport by water vapour generated through the interaction of hot magmatic fluids and the water table.  Taken together, especially with respect to the recent evolution at the summit, these findings indicate the importance of ground heating and thermal anomalies as a precursor for unrest of volcanic sites such as La Soufrière\deleted{ de Guadeloupe} which may be far more relevant than at caldera-type sites (e.g. Campi Flegrei or Nisyros) where CO$_2$ degassing is far more pervasive and heat loss through the ground is dominant.

%
\section*{Conclusions}

La Soufrière\deleted{ de Guadeloupe} is an andesitic \deleted{type} stratovolcano in the lesser Antilles arc with an extensive hydrothermal system that has undergone six phreatic/hydrothermal eruptions since 1635 C.E.  Here, we have concentrated on using thermal measurements to highlight the changes to the system over the past two decades which cover most of the current unrest since its onset in 1992.  Direct measurements were made of the temperature and mass flux at the key fumarolic emission sites and at numerous thermal springs linked to the hydrothermal activity of La Soufrière\deleted{ de Guadeloupe}.  The ground temperature at sites showing extensive thermal anomalies was determined from airborne thermal imagery.  From these and ancillary measurements for ambient conditions, we have deduced heat and mass fluxes as well as heat flux densities.  We have compared and discussed our measurements in light of historic data available in the literature.  Based on a reinterpretation of previously published data, we deduce that fumarolic output has proportionally decreased from 95\% of the total heat budget in 2010 to 78\% currently, whereas ground heating has increased from 4\% in 2010 to 21\% currently.  The present-day \replaced{convective and radiative heat fluxes in the summit area of 4.94 and \SI{0.74}{\MW}}{radiative and convective heat fluxes in the summit area of 0.74 and 4.94 MW}, respectively, have increased by an order of magnitude in the past decade which is \replaced{largely}{in large part} due to an increase in heated area and also, to a lesser degree, because of increased temperatures.  The total heat flux density presently estimated at \SI{403(24)}{\Wmsq} is greater than the 2010 estimate of \SI{326(69)}{\Wmsq}, and thus suggests an increase in thermal intensity at the summit, though these values are within the bounds of measurement uncertainty.  \replaced{These changes are explained partly by a spreading in fumarolic sites over the dome during the past decade but also, and more importantly, that ground thermal anomalies on the summit have propagated significantly in recent years.  }{These changes have occurred concurrently with continued and increasing rates of the opening of fractures on the dome as well as steady horizontal radial displacement of 3-10 mm/year for sites on the dome as measured with GNSS.  }
\added{Fractures on the dome along with steady horizontal radial displacements of 3--10 mm/year have been observed over the same period.  }
The thermal spring activity has changed little in 20 years although several of the thermal springs closest to the dome (GA, TA, BJ, PR) have shown since 2000 a steady linear increase of their temperature and heat flux rate. 

We find that, in terms of heat flux density (heat loss per unit area), La Soufrière\deleted{ de Guadeloupe} is amongst the most intense emitters of heat for volcanoes worldwide, and that its ranking has dramatically increased in recent years.
\replaced{With recent unrest events in mind, plus petrological evidence and geochemical analysis of magmatic fluids, we must consider that conditions with the potential to lead to phreatic/hydrothermal events currently exist at La Soufrière.}
         {Given recent unrest events, our assessments point to the presence of magma at or below the brittle-ductile transition releasing heat and fluids and likely subject to periodic refilling from the deeper parts of the magmatic system.}
\deleted{These results are coherent with recent petrological analysis of the volcano's last major magmatic eruption in 1530 C.E. by Pichavant et al. (2018) placing a shallow magma reservoir between a depth of 5 and 8 km as well as the interpretation of the geochemistry of emitted fluids by Villemant et al. (2014) and Moretti et al. (2020a,b) that infer a contribution of deep magmatic gases that trigger periodic transitory heating and pressurisation of the deep hydrothermal system near or slightly above the critical point of water at a depth of about 2-3 km below the summit.}
\deleted{Despite the current context of a lack of any evidence for the ascent of magma to very shallow depths, it is important to keep in mind in a hazard perspective, that larger, more frequent, or more intense transitory pulses of hot magmatic gases could exceed the buffering capacity of the hydrothermal system as well as the convective efficiency of heat transfer in a relatively and locally open non-sealed host-rock, bringing the overall system to critical conditions compatible with phreatic/hydrothermal eruptive unrest.}
Hence, La Soufrière\deleted{ de Guadeloupe} remains the subject of continued and enhanced surveillance and research strategies to better understand the origin of unrest and track its dynamic evolution.

\section*{Acknowledgements}
  The authors thank: the Assistant Editor, M.~R.~James and two anonymous authors for their constructive comments and suggestions; the OVSG-IPGP team for logistical support and help with data collection, Pierre Agrinier, and especially Gilbert Hammouya and Olivier Crispi for data collection before 2013;
  Pascal Allemand and IGN for DEMs and orthophotos;
  the Préfecture de Guadeloupe and the pilots of the Dragon 971 helicopter base in Guadeloupe (Sécurité Civile, Ministère de l'Intérieur) for providing helicopter support;
  the Parc National de Guadeloupe for assistance and authorisation of research and monitoring on La Soufrière;
  IPGP, INSU-CNRS through the Service National d'Observation en Volcanologie (SNOV), and the Ministère pour la Transition Écologique et Solidaire (MTES) for financial support.
  This work has been supported by the ANR DOMOSCAN, ANR DIAPHANE, the AO-IPGP 2018 project ``Depth to surface propagation of fluid-related anomalies at La Soufrière de Guadeloupe volcano (FWI): timing and implications for volcanic unrest'' (coord.: R. Moretti), and the European Union's through EUROVOLC (project No 731070).
  This study contributes to the IdEx Université de Paris ANR-18-IDEX-0001, is IPGP contribution number 4164 and is LabEx ClerVolc contribution number 426.
  MJH acknowledges funding via the INSU-CNRS project ``Assessing the role of hydrothermal alteration on volcanic hazards''.

\section*{Author Contributions}

DEJ collected and analysed the field data, prepared the figures and wrote the manuscript.  SM, RM, DG, JCK and VR also collected and analysed field data.  All authors contributed in the writing and discussion of the manuscript, and consented to its submission.

\bibliographystyle{spmpsci}
\bibliography{bibliography}

\begin{table}[ht]
  \caption{comparison of radiative and convective fluxes, and heat flux 
    density for the ground-heated sites around La Soufrière
    \deleted{de Guadeloupe} 
    from 2019 aerial imagery.  In the right-hand column, mean values are 
    given for the flux densities, and total values for all other quantities.}
  \label{tab:Table1}       
  \centering
  \begin{adjustbox}{width=\columnwidth, center}
    \input{Table1.tex}
  \end{adjustbox}
\end{table}

\begin{table}[ht]
  \caption{Comparison of the mass and heat flux estimates in 2010
    \citep{GAU:JVGR2016} and for 2020 (present study).  Values in parentheses
    are reworked fluxes which, for 2010, are based on the likely evolution of
    fumarole fluxes (see Fig.\:\ref{fig:long_time_series}) and the thermal
    springs data (Fig.\:\ref{fig:thermal_springs}) and for the present-day
    fumarole estimates are the MultiGAS traverse results scaled to the 
    Pitot-tube measurements.  Dashes indicate that no data was acquired at 
    that site.  *Encompasses RC and Matylis.}
  \label{tab:Table2}
  \centering
  \begin{adjustbox}{width=\columnwidth, center}
    \input{Table2.tex}
  \end{adjustbox}
\end{table}

%
%


\end{document}

%% file: Table1.tex
\begin{tabular}{lrrrrrr}
\toprule
{} & Summit & Matylis (lower) & Ravine Claire &  FTY0 &  FTY1 & Total/Mean$^\ast$ \\
\midrule
Rad. flux density, $q_{\mathrm{rad}}$/[W/m$^2$]   &     52 &              76 &            50 &    71 &    62 &             54.59 \\
Conv. flux density, $q_{\mathrm{conv}}$/[W/m$^2$] &    350 &             455 &           310 &   432 &   383 &            351.16 \\
Total heat flux density/[W/m$^2$]                 &    403 &             531 &           360 &   504 &   446 &            405.75 \\
Radiative flux, $Q_{\mathrm{rad}}$/[MW]           &   0.74 &            0.12 &          0.40 &  0.10 &  0.07 &              1.43 \\
Convective flux, $Q_{\mathrm{conv}}$/[MW]         &   4.94 &            0.74 &          2.49 &  0.63 &  0.42 &              9.23 \\
Total ($Q_\mathrm{rad} + Q_\mathrm{conv}$)/[MW]   &   5.67 &            0.87 &          2.89 &  0.74 &  0.49 &             10.66 \\
Heated area/[m$^2$]                               &  14070 &            1630 &          8010 &  1460 &  1100 &             26270 \\
\bottomrule
\end{tabular}

%% file: Table2.tex
\begin{tabular}{cc|cccc|cccc|cccc|c}
  \toprule
  \multirow{2}{*}{Flux} & \multirow{2}{*}{Year} & \multicolumn{4}{c|}{Fumaroles} & \multicolumn{4}{c|}{Ground thermal anomaly} & \multicolumn{4}{c|}{Hot springs} & Total \\
  &  & CS & G56 & TAS & Total & Summit & FTY & Flanks$^\ast$ & Total & GA & PR & TA & Total & \\
  \midrule
  \multirow{3}{1.0cm}{Mass [kg/s]} & \multirow{2}{*}{2010} &
  $19.5\pm4.0$ & \multirow{2}{*}{-} & $22.8\pm8.1$ & $42.3\pm12.1$ &
  \multirow{2}{*}{-} & \multirow{2}{*}{-} & \multirow{2}{*}{-} & \multirow{2}{*}{-} &
  2.5 & - & 1.5 & 4.5 &
  $46.8\pm12.1$ \\
  &  &  ($5.3\pm1.1$) &  &  ($6.2\pm2.2$) &  ($11.5\pm3.3$) &
  & & & &
  (2.51) & (0.44) & (1.61) & (4.56) & 
  ($16.06\pm3.3$) \\
  & 2020 & $4.9 \pm 0.5$ & ($2.1\pm0.8$) & ($2.8\pm1.1$) & ($9.8\pm2.5$)
         & -    & -    & -    & -
         & 3.35 & 0.44 & 1.61 & 5.40
         & ($15.2 \pm 2.5$) \\

  \midrule
  \multirow{3}{1.0cm}{Heat [MW]} & \multirow{2}{*}{2010} &
  $48.0\pm9.8$ & \multirow{2}{*}{-} & $56.1\pm19.9$ & $104.1\pm20.7$ &
  \multirow{2}{*}{$0.2\pm0.1$} & \multirow{2}{*}{$1.0\pm0.2$}
     & \multirow{2}{*}{-} & \multirow{2}{*}{$1.2\pm0.3$} &
  0.3 & - & 0.2 & 0.5 &
  $105.8\pm21.0$ \\
  &  & ($13.0\pm2.6$) &  & ($15.2\pm5.4$) & ($28.2\pm8.0$) &
  &  &  &  &
  (0.25) & (0.02) & (0.12) & (0.39) &
  ($29.8\pm8.3$) \\
  & 2020 & $15.3\pm1.5$ & ($5.5\pm2.2$) & ($7.5\pm3.0$) & ($28.3\pm6.8$)
         & $5.7\pm0.9$ & $0.6\pm0.1$ & $1.3\pm0.2$ & $7.6\pm1.1$
         & 0.36 & 0.01 & 0.19 & 0.56 
         & ($36.5\pm7.9$) \\
  \bottomrule
\end{tabular}